\documentclass[12pt]{iopart}
\pdfoutput=1
\usepackage{iopams}
\usepackage{cite,graphicx,multirow,dsfont,xcolor,float}
\usepackage[colorlinks]{hyperref}

\usepackage{makecell}

\newcommand{\1}{\mathds{1}}

\newcommand{\bA}{{\bf A}}
\newcommand{\bF}{{\bf F}}
\newcommand{\bW}{{\bf W}}
\newcommand{\bX}{{\bf X}}
\newcommand{\bbA}{\mathbb{A}}
\newcommand{\bbB}{\mathbb{B}}

\begin{document}

\title[Smallest eigenvalue density for complex Wishart-Laguerre ensemble]{Smallest eigenvalue density for regular or fixed-trace complex Wishart-Laguerre ensemble and entanglement in coupled kicked tops}

\author{Santosh Kumar \& Bharath Sambasivam \& Shashank Anand}

\address{Department of Physics, Shiv Nadar University, Gautam Buddha Nagar, Uttar Pradesh -- 201314, India}
\ead{skumar.physics@gmail.com}
\vspace{10pt}

\begin{abstract}
The statistical behaviour of the smallest eigenvalue has important implications for systems which can be modeled using a Wishart-Laguerre ensemble, the regular one or the fixed trace one. For example, the density of the smallest eigenvalue of the Wishart-Laguerre ensemble plays a crucial role in characterizing multiple channel telecommunication systems. Similarly, in the quantum entanglement problem, the smallest eigenvalue of the fixed trace ensemble carries information regarding the nature of entanglement. 

For real Wishart-Laguerre matrices, there exists an elegant recurrence scheme suggested by Edelman to directly obtain the exact expression for the smallest eigenvalue density. In the case of complex Wishart-Laguerre matrices, for finding exact and explicit expressions for the smallest eigenvalue density, existing results based on determinants become impractical when the determinants involve large-size matrices. In this work, we derive a recurrence scheme for the complex case which is analogous to that of Edelman's for the real case. This is used to obtain exact results for the smallest eigenvalue density for both the regular, and the fixed trace complex Wishart-Laguerre ensembles. We validate our analytical results using Monte Carlo simulations. We also study scaled Wishart-Laguerre ensemble and investigate its efficacy in approximating the fixed-trace ensemble. Eventually, we apply our result for the fixed-trace ensemble to investigate the behaviour of the smallest eigenvalue in the paradigmatic system of coupled kicked tops.

\end{abstract}

%
%
%
%
%

\section{Introduction}
\label{Sec1}

Wishart-Laguerre ensembles constitute an important class of random matrix ensembles~\cite{Mehta2004,Forrester2010} and have found diverse applications in the field of multivariate statistics~\cite{Anderson2003,Muirhead2005,James1964}, problems related to time series~\cite{Gnanadesikan1997,PGRAGS2002,VP2010}, low energy quantum chromodynamics~\cite{Verbaarschot1994,VW2000}, multiple-channel telecommunication~\cite{TV2004,FG1998,Telatar1999}, quantum entanglement~\cite{Lubkin1978,LP1988,Hall1998,Page1993,ZS2001,SZ2004,OSZ2010}, etc. Often the smallest eigenvalue distribution plays a crucial role in investigating the behaviour of the system under study. For instance, in the context of multiple input multiple output (MIMO) communication, the smallest eigenvalue of Wishart-Laguerre ensemble determines the minimum distance between the received vectors~\cite{Burel2002}, and also the lower bound on the channel capacity that eventually is used in antenna selection techniques~\cite{PL2006}. Similarly, the smallest eigenvalue density of fixed trace Wishart-Laguerre ensemble serves as an important metric for characterizing the entanglement in bipartite systems~\cite{MBL2008,Majumdar2011}. 

Matrices governed by Wishart distribution are parametrized by their size ($n$) and the degree of freedom ($m$)~~\cite{Anderson2003,Muirhead2005,James1964}; see section \ref{Sec2}. In this paper we use the term {\it regular} to mean unrestricted trace Wishart matrices with $m\ge n$. The smallest eigenvalue of Wishart matrices have been studied extensively, both for regular, and fixed trace scenarios. Moreover, finite-dimension, as well as large-dimension asymptotic cases have been explored. Our focus here is on the finite-dimensional scenario with the primary objective to obtain explicit expressions for the smallest eigenvalue density.

In the case of regular Wishart-Laguerre ensemble, for real matrices at finite $n,m$, Edelman, among other things, has provided a recursion-based scheme to obtain the smallest eigenvalue density~\cite{Edelman1989,Edelman1991}. For complex matrices, the result for the cumulative distribution of the smallest eigenvalue  goes back to Khatri, who worked out a determinant-based expression~\cite{Khatri1964}. Forrester and Hughes have given closed expressions for the density of the smallest and second-smallest eigenvalues in terms of Wronskian and Toeplitz determinants~\cite{FH1994}. Further generalizations applicable to a wider symmetry class of Wishart matrices have been considered in~\cite{Forrester1993,NF1998,Forrester1994}. Damgaard and Nishigaki have derived the probability distribution of the $k$th smallest eigenvalue of Dirac operator in the microscopic scaling limit for real, complex as well as quaternion cases and demonstrated the universality of the results~\cite{DN2001}. These eigenvalues have direct relationship with those of the Wishart-Laguerre ensemble. In~\cite{AGKWW2014} Akemann {\it et al.} have further explored the smallest eigenvalue distribution of real Wishart-Laguerre matrices and validated the universality in the microscopic limit for the correlated case also. Moreover, in a very recent work by Edelman, Guionnet, and P\'ech\'e~\cite{EGP2016}, the behaviour of the smallest eigenvalue coming from finite random matrices (including Wishart) based on non-Gaussian entries has been investigated.

For the fixed trace case, Chen, Liu and Zhou~\cite{CLZ2010} have derived the smallest eigenvalue density in terms of sum of Jack polynomials. Moreover, for the complex case this expression has been simplified to inverse Laplace transform of a certain determinant. In~\cite{AV2011}, for the real Wishart matrices, Edelman's recursive approach for the regular Wishart-Laguerre ensemble has been used by Akemann and Vivo to obtain the smallest eigenvalue density for the fixed trace Wishart-Laguerre ensemble.

For the complex case, the exact result for the smallest eigenvalue density is available in terms of determinants involving $n$-dimensional~\cite{Khatri1964,Burel2002,ZCW2009} or $\alpha$-dimensional matrices~\cite{FH1994,CLZ2010}, where $\alpha=m-n$ is the {\it rectangularity}. These results have been used for asymptotic analysis when $n\to\infty$ for $\alpha$ fixed and these correspond to eigenvalue distributions comprising Bessel kernel~\cite{FH1994,NDW1998,DN2001}. On the other hand, if both $n,\alpha\to\infty$, an analysis involving Fredholm determinant with Airy kernel is possible and that leads to the celebrated Tracy-Widom (TW) distribution~\cite{TW1993,TW1994a,TW1994b}. In~\cite{BF2003} the transition regime between the Bessel and Airy densities has also been investigated. While these asymptotic results give a wealth of information regarding the universal aspects, if one desires explicit expressions for the smallest eigenvalue density for large but finite $n,\alpha$ then these determinant based results turn out to be impractical, even with the availability of modern computational packages. We should remark that the smallest eigenvalue density has also been worked out for correlated Wishart-Laguerre ensembles, both for real and complex cases~\cite{ZCW2009,Forrester2007,WG2013}. However, the exact results are, again, in terms of determinants or Pfaffians. 

The iterative scheme provided by Edelman~\cite{Edelman1989,Edelman1991} is quite an effective and convenient way to calculate the density for the case of real matrices, and one can handle large values of dimensionality $n$ and rectangularity $\alpha$. For the complex Wishart-Laguerre ensemble, Forrester and Hughes have worked out an iterative scheme for the cumulative distribution of the smallest eigenvalue. However, to the best of our knowledge, an iterative scheme analogous to that of Edelman's, for direct evaluation of the probability density of the smallest eigenvalue has hitherto remained unavailable. In the present work, we derive the recurrence scheme that applies to the complex Wishart-Laguerre ensemble. These results involve an `exponential times polynomial' structure. Since the fixed trace ensemble is related to the regular Wishart-Laguerre ensemble via a Laplace transform, the structure of the smallest eigenvalue density in the latter leads to a very convenient calculation of density in the former case as well~\cite{AV2011}. Moreover, arbitrary moments of the smallest eigenvalue are also readily obtained. In addition, for the regular Wishart-Laguerre ensemble we also indicate a relation between the recurrence relation and the determinantal results of Forrester and Hughes~\cite{FH1994}, and explicitly demonstrate the equivalence between the two results for rectangularity $\alpha=0,1$. Similarly, for the fixed-trace scenario we prove the equivalence of the recursion-based expression and the result of Chen, Liu and Zhou~\cite{CLZ2010} based on the inverse Laplace transform of a determinant, again for $\alpha=0,1$.

Finally, we use the smallest eigenvalue density of the fixed trace ensemble to study entanglement formation in the paradigmatic system of coupled kicked tops. We should note that although the Floquet operator involved in this system belongs to the circular orthogonal ensemble (COE)~\cite{Mehta2004,Forrester2010}, the dynamically generated states are not random real vectors in the Schmidt basis~\cite{TMD2008}. Rather, they are complex, and hence, the results for the complex fixed trace Wishart-Laguerre ensemble are applicable. 

\section{Wishart-Laguerre ensemble}
\label{Sec2}
Consider complex matrices ${\bf A}$ of dimensions $n\times m$ taken from the density $\mathcal{P}_A(\bA)\propto \exp\left(-\tr \bA\bA^\dag\right)$. 
We assume without loss of generality that $n\leq m$. Then, the non-negative definite matrices $\bW=\bA\bA^\dag$ constitute the (regular) Wishart-Laguerre ensemble with the probability density
\begin{equation}
\mathcal{P}_W(\bW)\propto (\det \bW)^{m-n}\exp\left(-\tr \bW\right).
\end{equation}
The joint probability density of unordered eigenvalues $(\lambda_j\ge 0, j=1,...,n)$ of $\bW$ is given by~\cite{Mehta2004,Forrester2010}~\footnote{We should note that $m\times m$--dimensional matrices $\bA^\dag\bA$ share the eigenvalues $\lambda_1,...,\lambda_n$ of $\bA\bA^\dag$. The other $m-n$ eigenvalues are all zeros. The joint density in this case can be written by introducing delta-functions for these zero-eigenvalues and implementing proper symmetrization among all eigenvalues.}
\begin{equation}
\label{JPD_WL}
P(\lambda_1,...,\lambda_n)=C_{n,\alpha}\, \Delta^2_n(\{\lambda\}) \prod_{j=1}^n\lambda_j^{\alpha}\, e^{-\lambda_j},
\end{equation}
with $\alpha=m-n$, and 
\begin{equation}
C^{-1}_{n,\alpha}=\prod_{j=1}^n\Gamma(j+1)\Gamma(j+\alpha).
\end{equation}
Also, $\Delta_n(\{\lambda\})=\prod_{1\leq k<j\leq n}(\lambda_j-\lambda_k)$ is the Vandermonde determinant.
For this classical ensemble, exact expression for correlation functions of all orders are known~\cite{Mehta2004,Forrester2010}. For example, the first order marginal density (one-level density),
\begin{equation}
p(\lambda)=\int_0^\infty d\lambda_2\cdots \int_0^\infty d\lambda_n P(\lambda,\lambda_2,...,\lambda_n),
\end{equation}
is given by
\begin{eqnarray}
\nonumber
p(\lambda)&=&\frac{1}{n}e^{-\lambda}\lambda^\alpha \sum_{j=0}^{n-1}\frac{\Gamma(j+1)}{\Gamma(j+\alpha+1)}\left(L_j^{(\alpha)}(\lambda)\right)^2\\
&=&\frac{\Gamma(n)}{\Gamma(m)}e^{-\lambda}\lambda^\alpha [L_{n-1}^{(\alpha)}(\lambda)L_n^{(\alpha+1)}(\lambda)-L_n^{(\alpha)}(\lambda)L_{n-1}^{(\alpha+1)}(\lambda)].
\end{eqnarray}
Here $L_i^{(\gamma)}(\lambda)$ represents the associated Laguerre polynomials~\cite{Szego1975}.

We now focus on the statistics of the smallest eigenvalue of $\bW$. The probability density for the smallest eigenvalue can be calculated using the joint probability density~(\ref{JPD_WL}) as~\cite{Edelman1989,Edelman1991,FH1994}
\begin{equation}
\label{fWdef}
f(x)=n\int_x^\infty d\lambda_2\cdots \int_x^\infty d\lambda_n \,P(x,\lambda_2,...,\lambda_n).
\end{equation}
As shown in \ref{AppA}, this can be brought down to the form
\begin{equation}
\label{fW}
f(x)=c_{n,m}\, e^{-nx}x^\alpha g_{n,m}(x).
\end{equation}
Here the normalization factor $c_{n,m}$ is given by
\begin{equation}
c_{n,m}=\frac{1}{\Gamma(n)\Gamma(m)}\prod_{i=1}^{n-1}\frac{\Gamma(i+2)}{\Gamma(i+\alpha)},
\end{equation}
and $g_{n,m}(x)$ is obtained using the following recurrence scheme:
\begin{eqnarray*}
&&{\bf I.~} \mathrm{ Set ~} S_0=g_{n,m-1}(x), S_{-1}=0\\
&&{\bf II.~} \mathrm{ Iterate~ the~ following~ for~ }i=1 \mathrm{ ~to~ } n-1:\\
&&~~~~~~S_i=(x+m-i+1)S_{i-1}-\frac{x}{n-i}\frac{d S_{i-1}}{dx}\\
&&~~~~~~~~~~~+x\,(i-1)\frac{m-i}{n-i}S_{i-2}\\
&&{\bf III.~} \mathrm { ~Obtain~ } g_{n,m}(x)=S_{n-1}
\end{eqnarray*}
The initial case ($m=n$) is given by $g_{n,n}(x)=1$. Thus, starting from the square case, for which the result is simple ($f(x)=ne^{-nx}$), we can go up to any desired rectangularity by repeated application of the above recurrence scheme. We note that~(\ref{fW}) is of the form
\begin{equation}
\label{fW1}
f(x)=\sum_{j=\alpha+1}^{\alpha n+1}h_j x^{j-1} e^{-n x},
\end{equation}
where $h_j$ are $n,m$ dependent rational coefficients. The lower and upper limits of the summation in~(\ref{fW1}) are $\alpha+1$ and $\alpha n+1$, respectively. This is because the recurrence scheme applied for rectangularity $\alpha$ gives $g_{n,m}(x)$ as a polynomial of degree $\alpha(n-1)$, and there is the factor $x^\alpha$ in~(\ref{fW}). The coefficients $h_j$ can be extracted once the above recursion has been applied.

The above simple structure for the probability density gives easy access to the $\eta$--th moment of the smallest eigenvalue of the Wishart-Laguerre ensemble. We obtain
\begin{equation}
\label{xeta}
\langle x^\eta\rangle=\int_0^\infty x^\eta f(x)\,dx=\sum_{j=\alpha+1}^{\alpha n+1}\frac{h_j}{n^{j+\eta}}\Gamma(j+\eta).
\end{equation}
We would like to remark that this relationship holds not only for non-negative integer values of $\eta$, but also for any complex $\eta$ such that Re$(\eta)>-\alpha-1$. 

In~\ref{Codes} we provide simple Mathematica~\cite{Mathematica} codes that produce exact results for the density as well as the $\eta$--th moment for any desired value of $n, m$ by implementing the above results. 

\begin{figure}[h!]
\centering
\includegraphics[width=1\linewidth]{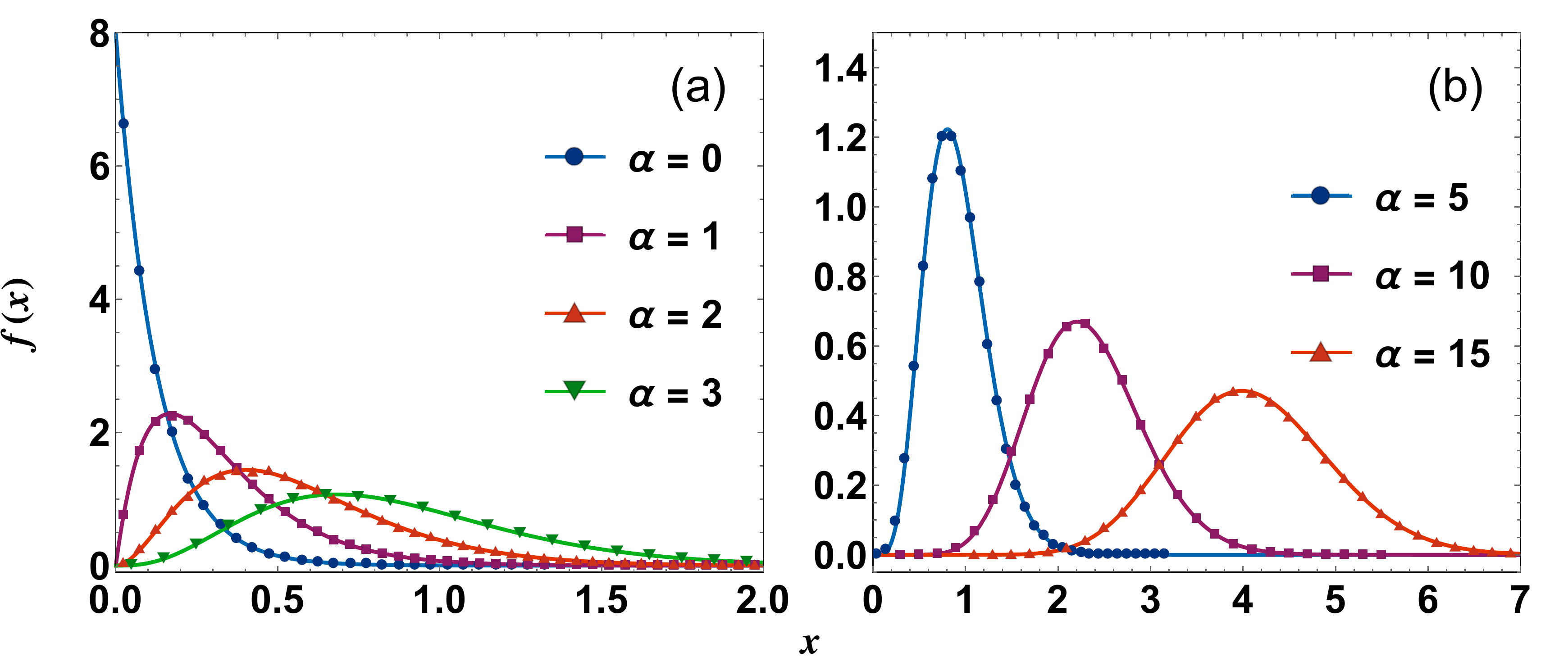}
\caption{\small Probability density of the smallest eigenvalue for the Wishart-Laguerre ensemble with (a) $n=8$, (b) $n=15$, and $\alpha$ values as indicated. The solid lines are analytical predictions based on~(\ref{fW}), while the symbols (filled- circles, squares, triangles) represent results of Monte Carlo simulations.}
\label{WLSev}
\end{figure}

In figure~\ref{WLSev} we consider the smallest eigenvalue density and compare the analytical results with Monte Carlo simulations using $10^5$ matrices for $n=8, 15$, and several $\alpha$ values. We find excellent agreement in all cases. 

Forrester and Hughes' result for the smallest eigenvalue density reads~\cite{FH1994}
\begin{equation}
\label{FH}
\fl
f(x)=(-1)^{\alpha(\alpha-1)/2} \frac{\Gamma(n+1)}{\Gamma(m)}e^{-n x}x^\alpha  \mathrm{det}\Big[ D_t^{\alpha+j-k-1}  L_{m-2}^{(3-\alpha)} (t) |_{t=-x}\Big]_{j,k=1,...,\alpha},
\end{equation}
where $D_t\equiv d/dt$. Comparing this result with~(\ref{fW}), we find that 
\begin{equation}
\label{equality}
\fl
g_{n,m}(x)=(-1)^{\alpha(\alpha-1)/2} \frac{\Gamma(n+1)}{\Gamma(m)\,c_{n,m}}  \mathrm{det}\Big[ D_t^{\alpha+j-k-1}  L_{m-2}^{(3-\alpha)} (t) |_{t=-x}\Big]_{j,k=1,...,\alpha}.
\end{equation}
Therefore, the recurrence scheme essentially leads to the evaluation of the above determinant, which otherwise becomes difficult to evaluate directly for large $\alpha$ values. Demonstrating the equality of the two sides in~(\ref{equality}) directly seems challenging for arbitrary $n,m$, if at all feasible. However, as shown below, for $\alpha=1$, we find that $g_{n,m}(x)$ does lead to the associated Laguerre polynomial as evaluated by the determinantal expression. Before analyzing the results of $\alpha=1$, we also consider $\alpha=0$, which corresponds to the {\it square} case.

\subsection{Results for $\alpha=0$}

In this case $g_{nm}=1$ and in the expression~(\ref{fW1}), there is just one term in the sum, viz. $j=1$. The corresponding value of the coefficient $h_j$ is $n$. Thus, the smallest eigenvalue density reads
\begin{equation}
\label{fa0}
f(x)=n e^{-n x}.
\end{equation}
Also, the moment-expression is given by
\begin{equation}
\langle x^\eta\rangle=\frac{\Gamma(\eta+1)}{n^\eta}.
\end{equation}
These expressions agree with those derived in~\cite{Edelman1989,FH1994}, as they should. We note that~(\ref{FH}) leads to the correct density, as the determinant part has to be taken as 1 for $\alpha=0$.

\subsection{Results for $\alpha=1$}

This is a nontrivial case. As shown in~\ref{EqProof}, in this case, $S_i$ in the recurrence relation can be identified with $\Gamma(i+1)L_i^{n-i+1}(x)$. Consequently, $g_{n,n+1}(x)=\Gamma(n)L_{n-1}^{(2)}(-x)$. Also $c_{n,n+1}=1/\Gamma(n)$, which leads to the smallest eigenvalue expression
\begin{equation}
\label{fa1}
f(x)=e^{-n x}x L_{n-1}^{(2)}(-x).
\end{equation}
This agrees with~(\ref{FH}) when evaluated for $\alpha=1$.
The use of the expansion of the Laguerre polynomial~\cite{Szego1975} leads to the coefficient $h_j$ in~(\ref{fW1}) as
\begin{equation}
\label{a1hj}
h_j=\frac{\Gamma(n+2)}{\Gamma(n-j+2)\Gamma(j+1)\Gamma(j-1)},~~~j=2,3,...,n+1.
\end{equation}
The $\eta$--th moment follows as
\begin{equation}
\langle x^\eta\rangle=\Gamma(n+2)\sum_{j=2}^{n+1}\frac{\Gamma(j+\eta)}{n^{j+\eta}\Gamma(n-j+2)\Gamma(j+1)\Gamma(j-1)}.
\end{equation}

\section{Fixed trace Wishart-Laguerre ensemble}
\label{Sec3}
\begin{figure}[h!]
\centering
\includegraphics[width=0.5\linewidth]{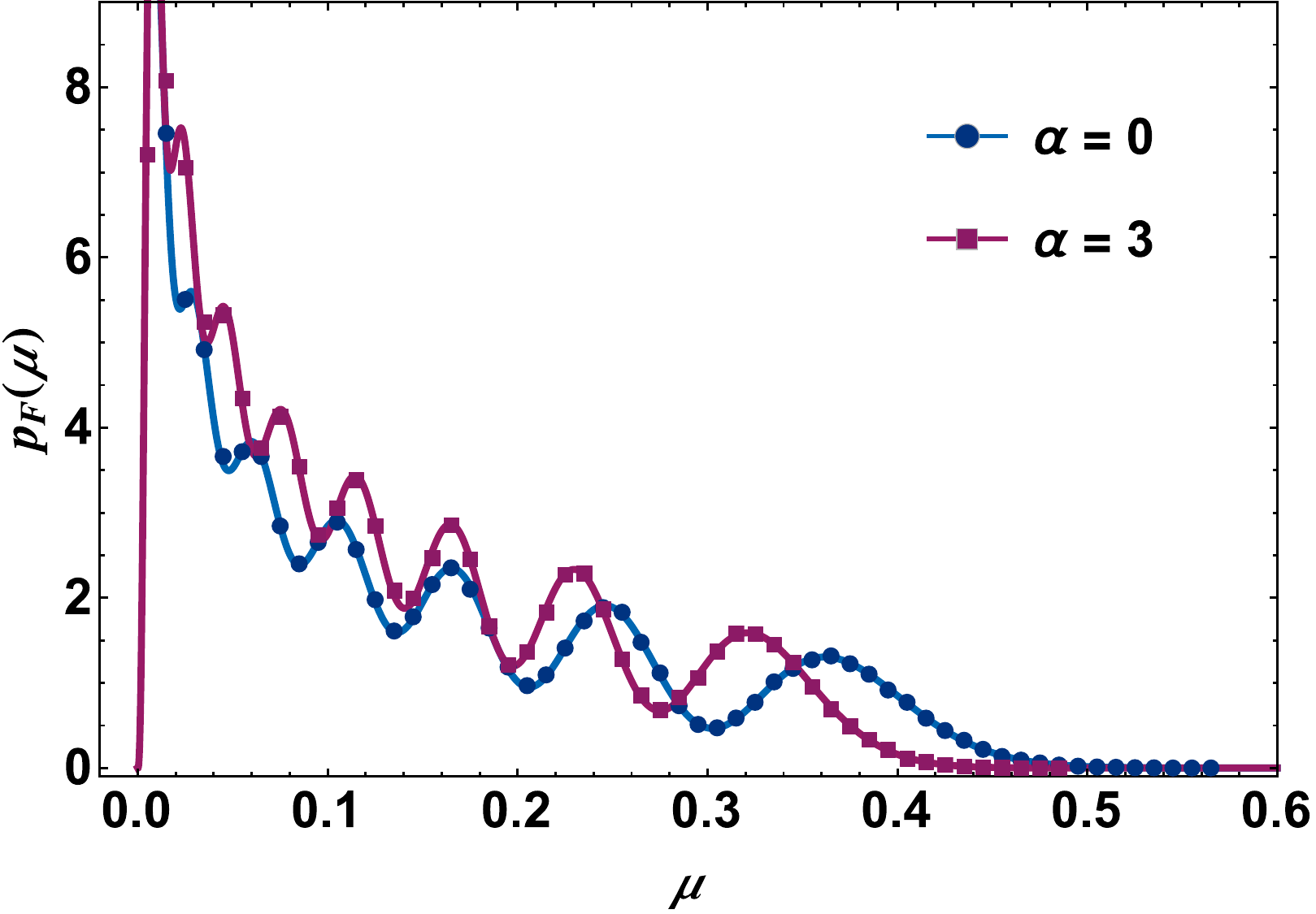}\\
\caption{\small Marginal density for fixed trace Wishart-Laguerre ensemble using~(\ref{PF1}) with $n=8$, and $\alpha$ values as indicated. The solid lines are analytical predictions based on (\ref{PF1}), and the symbols correspond to Monte Carlo results.}
\label{FTWL}
\end{figure}

Fixed trace ensembles constitute a special class of random matrices and can take care of system dependent constraints. For the Wishart-Laguerre case, the corresponding fixed trace ensemble arises naturally in the quantum entanglement problem in bipartite systems~\cite{LP1988,Hall1998,ZS2001}. With the trace value fixed at unity, it models the reduced density matrix; see section~\ref{EntBi}. Using the Wishart matrices $\bW$ from the preceding section, the fixed trace ensemble can be realized by considering the matrices $\bF=\bW/\tr\bW$~\cite{ZS2001,OSZ2010}. The corresponding probability density is
\begin{equation}
\label{PFT}
\mathcal{P}_F(\bF)\propto (\det \bF)^\alpha\, \delta(\tr\bF-1).
\end{equation}
The joint density of unordered eigenvalues $(0\leq \mu_j\leq 1; j=1,...,n)$ of $\bF$ is obtained as~\cite{LP1988,Hall1998,ZS2001}
\begin{equation}
\label{JPD_FTWL}
P_F(\mu_1,...,\mu_n)=C^F_{n,\alpha}\, \delta\left(\sum_{i=1}^n\mu_i-1\right)\Delta_n^2(\{\mu\}) \prod_{j=1}^n\mu_j^{\alpha},
\end{equation}
where $C^F_{n,\alpha}=\Gamma(nm)\,C_{n,\alpha}$~\cite{KP2011}. The corresponding marginal density has been derived in~\cite{ATK2009} as a single sum over hypergeometric $_5F_4$, and as a double sum over {\it polynomials} in~\cite{Vivo2010}. In~\cite{KP2011} it has been given as a single sum over the Gauss hypergeometric function ($_2F_1$):
\begin{eqnarray}
\label{PF1}
\nonumber
&&p_F(\mu)=\sum_{i=0}^{n-1} K_i\, \mu^{i+\alpha}(1-\mu)^{-i+n m-\alpha-2}\\
&&\times\left(n\,\mathcal{F}^{1-n,i-nm+\alpha+1}_{\alpha+1}-(n-i-1)\mathcal{F}^{-n,i-nm+\alpha+1}_{\alpha+1}\right).
\end{eqnarray}
Here we used the notation $\mathcal{F}^{a,b}_{c}:=\,_2F_1(a,b;c;\frac{\mu}{1-\mu})/\Gamma(c)$. Also, the coefficient $K_i$ is given by
\begin{equation}
K_i=\frac{(-1)^i\Gamma(m+1)\Gamma(nm)}{n\Gamma(i+1)\Gamma(n-i)\Gamma(i+\alpha+2)\Gamma(nm-\alpha-i-1)}.
\end{equation}
Figure~\ref{FTWL} shows the comparison between analytical and Monte Carlo results for the marginal density of the fixed trace ensemble. We find excellent agreement.

\begin{figure}[ht!]
\centering
\includegraphics[width=1\linewidth]{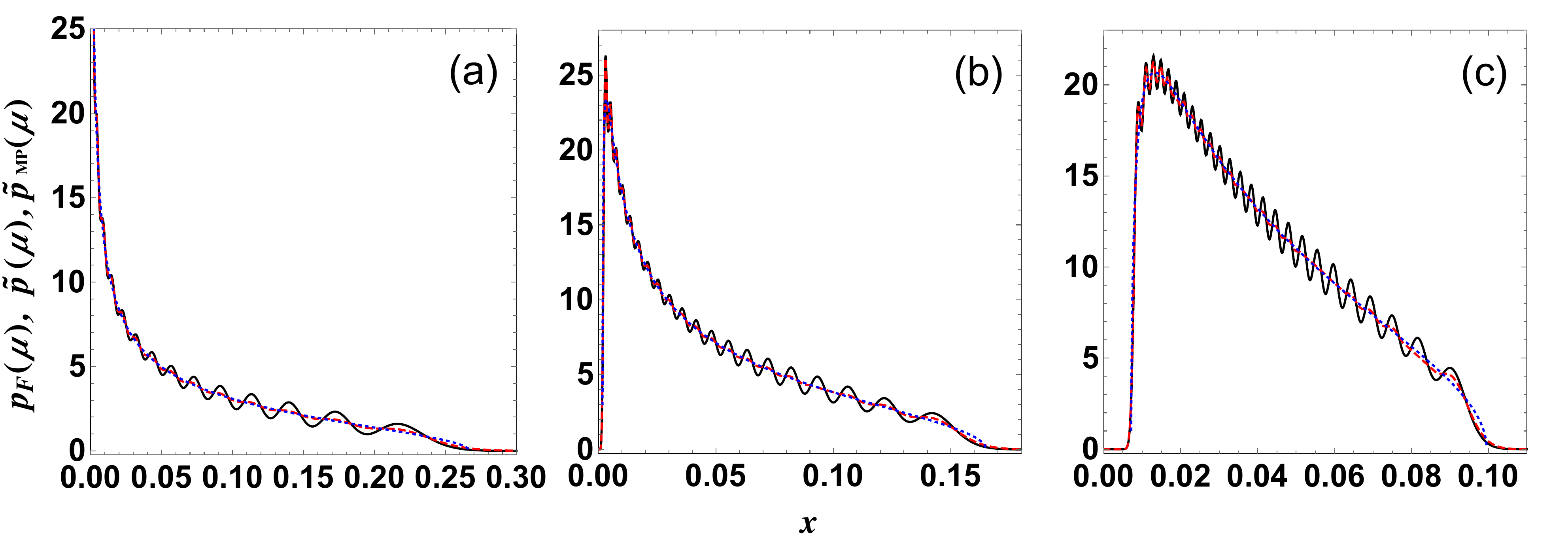}\\
\caption{\small Marginal density of eigenvalues for the fixed trace Wishart-Laguerre ensemble: Comparison between exact (solid black), scaled (dashed red), scaled Mar\v{c}enko-Pastur (dotted blue) as given by (\ref{PF1}), (\ref{PS1}) and (\ref{PS2}), respectively. (a) $n=m=15$, (b) $n=20, m=30$, (c) $n=25, m=75$.}
\label{ScWis}
\end{figure}

Using Selberg's integral and its properties~\cite{Mehta2004,Forrester2010}, it can be shown that both the average and the variance of the trace for the regular Wishart-Laguerre ensemble is $mn$. Therefore, if we consider the ensemble of matrices $\bW/mn$, the corresponding eigenvalues are $1/mn$ times the eigenvalues of $\bW$. Moreover, while individually these scaled matrices may not have trace one, on average, it is one. In addition, the variance of trace for this scaled ensemble is $1/mn$, which becomes negligible for large $n,m$. Therefore, it is expected that this scaled ensemble will approximate the behaviour of the fixed trace ensemble. For instance, the marginal density for this scaled ensemble,
\begin{equation}
\label{PS1}
\widetilde{p}(\mu)=m n\, p(mn \mu),
\end{equation} 
should serve as an approximation to $p_F(\mu)$. We can also use Mar\v{c}enko-Pastur density~\cite{MP1967} to write down an expression for $\widetilde{p}(\mu)$ valid for large $n,m$:
\begin{equation}
\label{PS2}
\widetilde{p}_{\mathrm{MP}}(\mu)=\frac{m}{2\pi}\frac{\sqrt{(\mu_{+}-\mu)(\mu-\mu_{-})}}{\mu};~~~\mu_{\pm}=\frac{(1\pm\sqrt{n/m})}{n}.
\end{equation}
This relation of the fixed trace ensemble with the scaled ensemble has been used in~\cite{Znidaric2007,BL2002,BL2004,SZ2004}. In figure~\ref{ScWis} we plot the exact one-eigenvalue density~(\ref{PF1}) for the fixed trace ensemble, as well as the densities~(\ref{PS1}),~(\ref{PS2}) based on the scaled ensemble. We find that while the density obtained using the scaled ensemble is not able to capture the oscillations, it does capture the overall shape of the density quite well.

\begin{figure}[t!]
\centering
\includegraphics[width=1\linewidth]{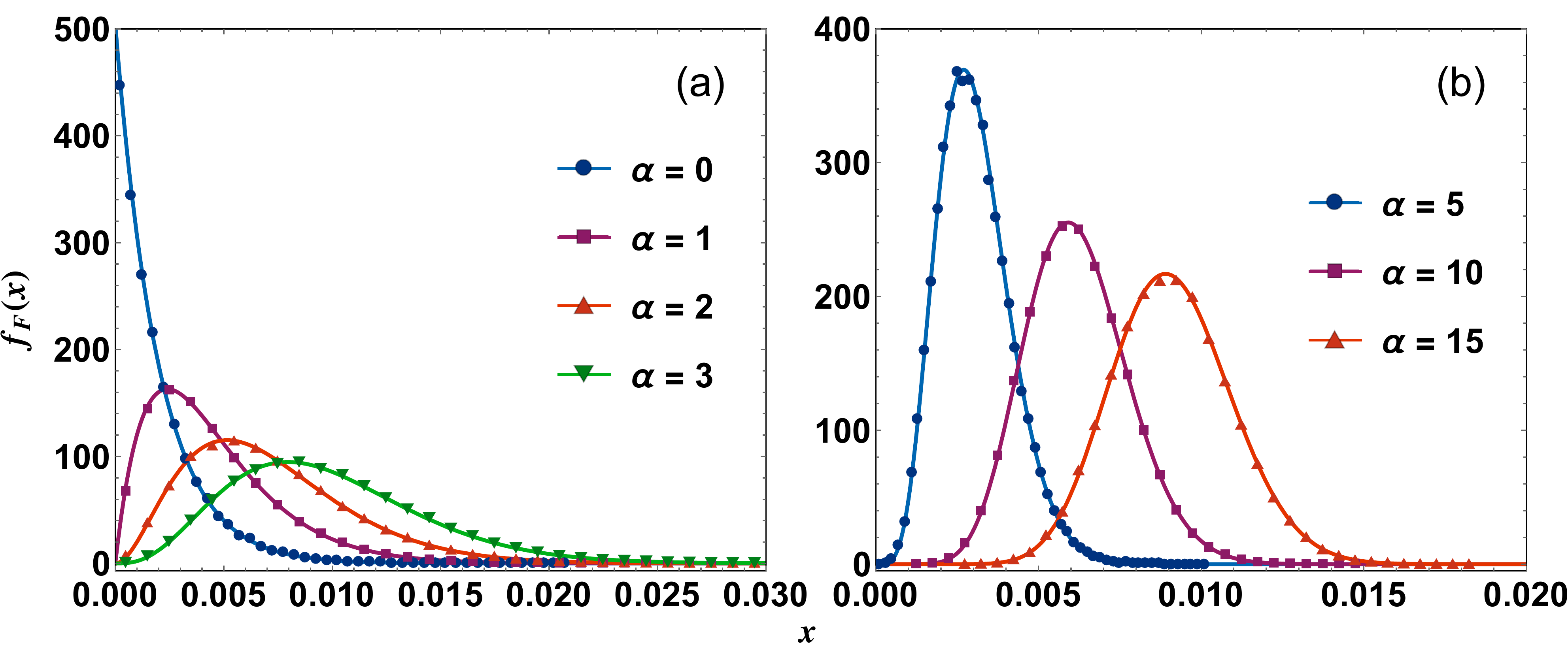}\\
\caption{\small Probability density of the smallest eigenvalue for the fixed trace Wishart-Laguerre ensemble with (a) $n=8$, (b) $n=15$, and $\alpha$ values as indicated. The solid lines are analytical predictions based on~(\ref{fFW}), while the symbols (filled- circles, squares, triangles) represent results of Monte Carlo simulations.}
\label{FTWLSev}
\end{figure}

The exact result for the density of the smallest eigenvalue for the fixed trace ensemble can be obtained using ~(\ref{fW}),~(\ref{fW1}), and the Laplace-inversion result 
\begin{equation}
\mathcal{L}^{-1}\{s^{-a} e^{-nsx}\}(t=1)=(1-n x)^{a-1}\Theta (1-n x)/\Gamma (a),
\end{equation}
with $\Theta(z)$ being the Heaviside-theta function. We have
\begin{eqnarray}
\label{fFW}
\nonumber
&f_F(x)=\Gamma(nm)\mathcal{L}^{-1}\{s^{1-nm}f(sx)\}(t=1)\\
&=\Gamma(nm)\sum_{j=\alpha+1}^{\alpha n+1}h_j\frac{(1-n x)^{n m-j-1}x^{j-1}}{\Gamma(nm-j)}\Theta(1-nx).
\end{eqnarray}
The prefactor $\Gamma(nm)$ comes from the ratio of normalizations, viz. $C^F_{n,\alpha}/C_{n,\alpha}$. In~\cite{AV2011} a similar strategy has been used for the real case. In figure~\ref{FTWLSev} we show the comparison between the analytical prediction and the numerical simulation for the smallest eigenvalue density. They are in excellent agreement.

\begin{figure}[h!]
\centering
\includegraphics[width=01\linewidth]{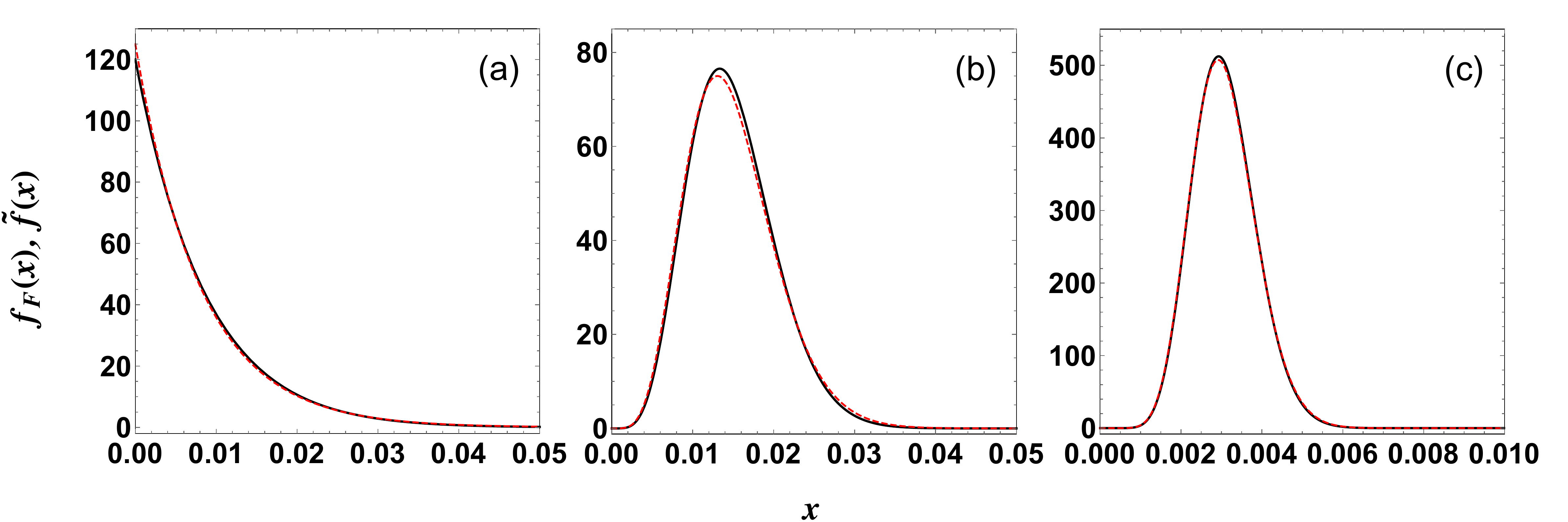}
\caption{\small Comparison between the exact ($f_F(x)$: solid black) and approximate ($\widetilde{f}(x)$: dashed red) probability densities for the smallest eigenvalue of the fixed trace Wishart-Laguerre ensemble, as given by (\ref{fFW}) and (\ref{AppfFW}), respectively. The parameter values used are (a) $n=m=5$, (b) $n=8, m=13$, and (c) $n=20, m=30$.}
\label{Approx}
\end{figure}

The idea of using scaled Wishart-Laguerre ensemble, as discussed above, can be applied here as well. Therefore, an approximate density for the smallest eigenvalue can be written using~(\ref{fW}) as 
\begin{equation}
\widetilde{f}(x)=m n f( m n x).
\label{AppfFW}
\end{equation}
In figure~\ref{Approx} we compare this approximation with the exact result given by~(\ref{fFW}). The approximation works pretty well. This approximate relation between the two densities is also the reason behind the very similar shapes of the curves in figures~\ref{WLSev} and \ref{FTWLSev}, respectively.

We also find that, using the first equality in~(\ref{fFW}), it follows that the $\eta$--th moment of the smallest eigenvalue for the fixed trace ensemble is related to that of the regular Wishart-Laguerre ensemble as
\begin{equation}
\langle x^\eta \rangle_F=\frac{\Gamma(nm)}{\Gamma(nm+\eta)}\langle x^\eta\rangle.
\end{equation}
This, similar to~(\ref{xeta}), holds for Re$(\eta)>-\alpha-1$.

Mathematica~\cite{Mathematica} codes to obtain explicit results for the above smallest eigenvalue density of the fixed trace ensemble, as well as the moments are given in~\ref{Codes}. 

Similar to the unrestricted trace case, we discuss below the cases $\alpha=0,1$ for the fixed trace scenario.
\subsection{Results for $\alpha=0$}

For $\alpha=0$ we just have one term in the series (\ref{fFW}), and $h_1=n$. Therefore, we arrive at
\begin{equation}
\label{fFa0}
f_F(x)=n (n^2-1)(1-n x)^{n^2-2}\,\Theta(1-nx).
\end{equation}
Also, the expression for the $\eta$-th moment is given by
\begin{equation}
\langle x^\eta\rangle_F=\frac{\Gamma(\eta+1)\Gamma(n^2)}{n^\eta\Gamma(n^2+\eta)}.
\end{equation}
These expressions are in agreement with those obtained in~\cite{MBL2008,Majumdar2011}.
\subsection{Results for $\alpha=1$}

In this case use of the result (\ref{a1hj}) for $h_j$ in (\ref{fFW}) leads to the smallest eigenvalue density expression
\begin{equation}
\label{fFa1}
\fl
f_F(x)=\Gamma(n^2+n)\Gamma(n+2)\sum_{j=2}^{n+1}\frac{(1-nx)^{n^2+n-j-1}x^{j-1}}{\Gamma(n-j+2)\Gamma(j+1)\Gamma(j-1)\Gamma(n^2+n-j)}\Theta(1-nx).
\end{equation}
Also, the $\eta$--th moment follows as
\begin{equation}
\langle x^\eta\rangle_F=\frac{\Gamma(n^2+n)\Gamma(n+2)}{\Gamma(n^2+n+\eta)}\sum_{j=2}^{n+1}\frac{\Gamma(j+\eta)}{n^{j+\eta}\Gamma(n-j+2)\Gamma(j+1)\Gamma(j-1)}.
\end{equation}
Chen, Liu and Zhou have provided the result for the cumulative distribution~\footnote{More appropriately, the survival function or the reliability function.} of the smallest eigenvalue for the complex case in terms of an inverse-Laplace transform involving a determinant~\cite{CLZ2010}:
\begin{equation}
\label{CLZ}
\fl
Q(x)=\Gamma(m n)x^{mn-1}\mathcal{L}^{-1}\Big\{s^{-m n}\det[L_{n+j-k}^{(k)}(-s)]_{j,k=0,...,\alpha-1} \Big\} \left(\frac{1-n x}{x}\right);~~0<x\le \frac{1}{n}.
\end{equation}
We set $\alpha=1$ in this expression and use the expansion for associated Laguerre polynomial~\cite{Szego1975}, later on. The inverse Laplace transform can then be explicitly performed leading us to
\begin{equation}
\fl
Q(x)=\Gamma(n+1)\Gamma(n^2+n) \sum_{j=0}^n \frac{x^j (1-n x)^{n^2+n-j-1}}{\Gamma^2(j+1)\Gamma(n-j+1)\Gamma(n^2+n-j)}.
\end{equation}
The probability density of the smallest eigenvalue follows upon using $f_F(x)=-dQ(x)/dx$ as
\begin{eqnarray}
\nonumber
\fl
f_F(x)=\Gamma(n+1)\Gamma(n^2+n)\sum_{j=0}^n \frac{n x^j (1-n x)^{n^2+n-j-2}}{\Gamma^2(j+1)\Gamma(n-j+1)\Gamma(n^2+n-j-1)}\\
-\Gamma(n+1)\Gamma(n^2+n)\sum_{j=1}^{n+1} \frac{x^{j-1}(1-nx)^{n^2+n-j-1}}{\Gamma(j)\Gamma(j+1)\Gamma(n-j+1)\Gamma(n^2+n-j)}.
\end{eqnarray}
In the second term we start the sum from $j=1$ as $j=0$ term is zero because of the diverging gamma function $\Gamma(j)$ in the denominator. Moreover, we have added a term $j=n+1$ which, again, is zero because of the diverging $\Gamma(n-j+1)$ in the denominator. Next, we consider $j\rightarrow j-1$ in the summand of the first term, and hence the sum runs from $j=1$ to $n+1$. The two terms can then be combined to yield (\ref{fFa1}) by noticing that the $n=1$ term is zero. We note that (\ref{CLZ}) also produces the correct result for $\alpha=0$ if the determinant value in this case is interpreted as 1.


\section{Large $n,\alpha$ evaluations and comparison with Tracy-Widom density}

The universality aspects of the regular Wishart-Laguerre ensemble have been explored in several notable works~\cite{Forrester1993,Forrester1994,TW1994b,NDW1998,DN2001,FS2010,EGP2016,AGKWW2014,TW1993,TW1994a,TV2010,KC2010,WKG2015,PS2016,B2016,NF1998}. For the fixed trace case, the local statistical properties of the eigenvalues have been studied in~\cite{CLZ2010,LZ2011}. In particular, it has been shown that the fixed trace and the regular Wishart-Laguerre ensembles share identical universal behaviour for large $n$ at the hard edge, in the bulk and at the soft edge for $\alpha$ fixed~\cite{LZ2011}. 

For the complex Wishart-Laguerre ensemble, in the square case ($\alpha=0$), the smallest eigenvalue scaled by $n$ gives rise to an exponential density~\cite{Edelman1989,Edelman1991,MBL2008}. Interestingly, this is true for all $n$ in this case, as evident from (\ref{fa0}). For large $n$ it has been shown that this result holds even if the matrices $\bW$ are constructed from non-Gaussian $\bA$~\cite{TV2010} ({\it cf.} section~\ref{Sec2}) with certain considerations. For the fixed trace case, in view of its connection to the scaled Wishart-Laguerre ensemble~(\ref{fFW}), as discussed in section~\ref{Sec3}, the smallest eigenvalue has to be scaled by $n^3$ to obtain the exponential density~\cite{MBL2008,Majumdar2011}. This can be easily verified to be true from~(\ref{fFa0}). Furthermore, very recently, $1/n$ corrections to the scaled smallest eigenvalue has been worked our for close to square cases~\cite{EGP2016,PS2016,B2016}. 
 \begin{figure}[ht!]
\centering
\includegraphics[width=1\linewidth]{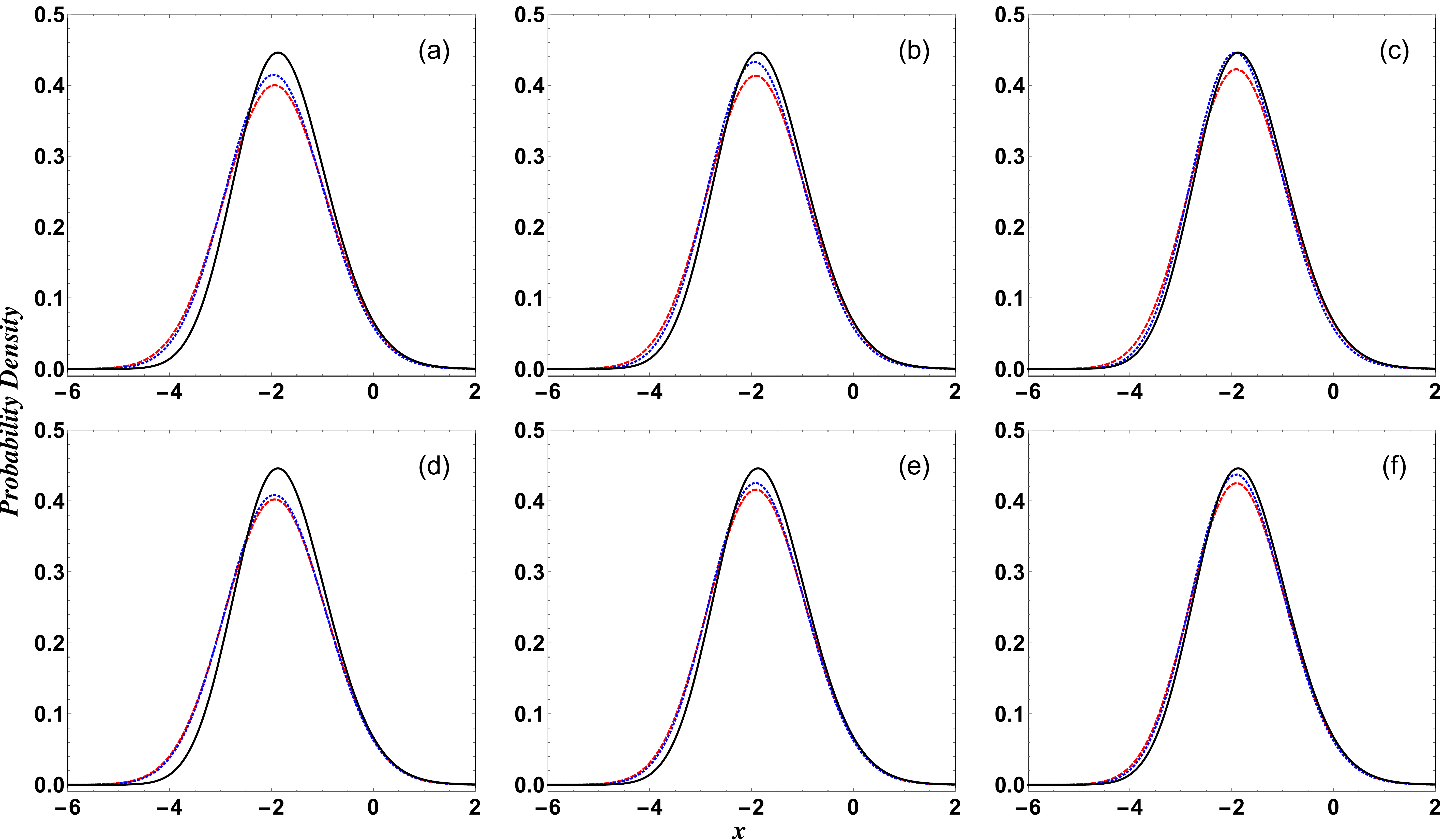}\\
\caption{\small Comparison of Tracy-Widom density (solid black) with densities $-\sigma f(\sigma x+\eta)$ (dashed red) and $-(\sigma/mn) f_F((\sigma x+\eta)/mn)$ (dotted blue) for (a) $n=25,m=125$, (b) $n=25,m=225$, (c) $n=25,m=425$, (d) $n=50,m=150$, (e) $n=50,m=250$, and (f) $n=50,m=450$. It should be noted that the rectangularity $\alpha$ varies as 100 for (a), (d); 200 for (b), (e); and 400 for (c), (f). Also, the aspect ratio $n/m$ is 1/5 for (a), (e), and 1/9 for (b), (f), respectively.}
\label{TW}
\end{figure}

For the rectangular case, Feldheim and Sodin~\cite{FS2010} have shown that, in the limit $m\rightarrow \infty$, $n\rightarrow \infty$ with $n/m$ bounded away from 1, the shifted and scaled smallest eigenvalue, $(\lambda_{\rm min}-\eta)/\sigma$, leads to the Tracy-Widom density~\cite{TW1993,TW1994a}. Here $\eta=(n^{1/2}-m^{1/2})^2$ and $\sigma=(n^{1/2}-m^{1/2})(n^{-1/2}-m^{-1/2})^{1/3}<0$. The convergence, however, is slower when $\alpha=m-n$ is close to 0. This can be attributed to the fact that the hard-edge behaviour is prevalent unless $\alpha$ is large~\cite{WKG2015}. We should also mention that the Tracy-Widom density captures the largest typical fluctuations of the smallest eigenvalue, while the larger atypical fluctuations are described by large deviation results, as derived in~\cite{KC2010} by Katzav and Castillo. 

As a consequence of identical universal behaviour of spectra of the regular and fixed-trace ensembles~\cite{LZ2011}, the Tracy-Widom density is also expected in the case of fixed trace ensemble. The proper scaling in this case can be inferred from the connection with the scaled Wishart-Laguerre ensemble, as discussed in Sec. 3. This implies that the density of $(m n \mu_{\rm min}-\eta)/\sigma$ will converge to the Tracy-Widom result. 

The recursion scheme given in section~\ref{Sec2} enables us to work out the exact results for the smallest eigenvalue density for large values of $n$ and $\alpha$ and hence to explore the above limit. In view of the scaling and shift indicated above, $-\sigma f(\sigma x+\eta)$ and $-(\sigma/mn) f_F((\sigma x+\eta)/mn)$ should coincide with the Tracy-Widom density of the unitarily-invariant class. We examine this in figure~\ref{TW}. We can see that as the rectangularity $\alpha$ increases the agreement improves for both $n=25$ and $n=50$. This is because the hard-edge effect is diminished with increasing $\alpha$. We also find that for a given aspect ratio $n/m<1$, as expected, the agreement is better for larger $n$.


\section{Entanglement in bipartite systems}
\label{EntBi}

Consider a bipartite partition of an $N_1 N_2$--dimensional Hilbert space $\mathcal{H}^{(N_1N_2)}$ consisting of subsystems $\bbA$ and $\bbB$, which belong to Hilbert spaces $\mathcal{H}_\bbA^{(N_1)}$ and $\mathcal{H}_\bbB^{(N_2)}$, respectively, such that $\mathcal{H}^{(N_1N_2)}=\mathcal{H}_\bbA^{(N_1)}\otimes \mathcal{H}_\bbB^{(N_2)}$. A general state $| \psi \rangle$ of $\mathcal{H}^{(N_1N_2)}$ is given in terms of the orthonormal states $| i^\bbA \rangle$ of $\mathcal{H}_\bbA^{(N_1)}$, and $| \alpha^\bbB \rangle$ of $\mathcal{H}_\bbB^{(N_2)}$ as
\begin{equation}
|\psi\rangle=\sum_{i=1}^{N_1}\sum_{\alpha=1}^{N_2}x_{i,\alpha} |i^\bbA\rangle \otimes |\alpha^\bbB\rangle,
\end{equation}
where $x_{i,\alpha}$ are complex coefficients, such that $\langle\psi|\psi\rangle=\sum_{i=1}^{N_1}\sum_{\alpha=1}^{N_2} |x_{i,\alpha}|^2=1$.
The density matrix for the composite system, considering a pure state scenario, is given by
  \begin{equation}
\rho=|\psi \rangle\langle \psi |=\sum_{i,j=1}^{N_1}\sum_{\alpha,\beta=1}^{N_2} x_{i,\alpha}x_{j,\beta}^{*}|i^\bbA\rangle \langle j^\bbA|\otimes |\alpha^\bbB\rangle\langle \beta^\bbB|,
\end{equation}
with $\tr[\rho]=1$. The reduced density matrix for subsystem, say $\bbA$, can be obtained by tracing out the subsystem $\bbB$ as
\begin{equation}
\label{rdmA}
\rho_\bbA=\sum_{\alpha', \beta'=1}^{N_2}\langle \alpha'|\rho|\beta'\rangle=\sum_{i,j=1}^{N_1}\mathcal{F}_{i,j}|i^\bbA\rangle\langle j^\bbA|,
\end{equation}
where $\mathcal{F}_{i,j}=\sum_{\alpha=1}^{N_2} x_{i,\alpha}x_{j,\alpha}^*$ can be viewed as the matrix elements of some $N_1\times N_1$--dimensional matrix $\bF=\bX\bX^\dag$. Here $\bX$ is a rectangular matrix of dimension $N_1\times N_2$ that has $x_{i,\alpha}$ as its elements. In the eigenbasis of $\bF$,~(\ref{rdmA}) can be written as
\begin{equation}
\rho_\bbA=\sum_{i=1}^{N_1}\mu_i |\mu_i^\bbA\rangle \langle \mu_i^\bbA|.
\end{equation}
The eigenvalues $\mu_i$ of $\bF$ are referred to as the Schmidt eigenvalues or Schmidt numbers. Due to the trace condition, they satisfy
\begin{equation}
\sum_{i=1}^{N_1} \mu_i=\tr\bX\bX^\dag=\tr \bF=1.
\end{equation} 
Suppose $N_1\le N_2$. Now, if we sample all normalized density matrices with equal probabilities, i.e., if we choose the coefficients $x_{i,\alpha}$ randomly using the Hilbert-Schmidt density $\mathcal{P}_X(\bX)\propto \delta(\tr \bX\bX^\dag-1)$, then $\bF$ defined here is statistically equivalent to the $\bF$ defined in~(\ref{PFT}), and the statistics of the Schmidt eigenvalues are described exactly by the joint eigenvalue density of the fixed trace Wishart-Laguerre ensemble~(\ref{JPD_FTWL}), with $N_1=n, N_2=m$~\cite{SZ2004,OSZ2010}. It should be noted that the reduced density matrix for the subsystem $\bbB$ will correspond to the matrix $\bX^\dag\bX$, which will share the eigenvalues $\mu_1,...,\mu_n$, and will have the rest of its $m-n$ eigenvalues as zero. As such, it carries the same amount of {\it information} as $\bF$.

The Schmidt eigenvalues can be used to study various entanglement measures such as von-Neumann entropy, Renyi entropies, concurrence, purity etc. As a consequence, fixed trace Wishart-Laguerre ensemble has been extensively used to model the reduced density matrices arising in the study of entanglement formation in bipartite systems~\cite{Lubkin1978,LP1988,OSZ2010,MBL2008,Majumdar2011,BL2002,BL2004,KAT2013,Page1993,Facchi2008,Giraud2007,Znidaric2007,NMV2010,NMV2011,ATK2009,CLZ2010,AV2011,Vivo2010,KP2011,LZ2011,VPO2016}. These works have explored several aspects such as moments and distributions of Schmidt eigenvalues and entanglement measures.

The density of the minimum eigenvalue in the present context not only sheds light on the nature of the entanglement, but also provides  important information about the degree to which the effective dimension of the Hilbert space of the smaller subsystem can be reduced~\cite{MBL2008,Majumdar2011}. The smallest eigenvalue assumes values from 0 to $1/n$. In the extreme case of $1/n$, it follows from the trace constraint $\sum_{i=1}^n \mu_i=1$, that all the eigenvalues must have the same value $1/n$. Consequently, the von-Neumann entropy, $-\sum_{i=1}^n \mu_i \ln \mu_i$, assumes its maximum value $\ln n$, thereby making the corresponding state {\it maximally} entangled. In the other extreme of the smallest eigenvalue being 0 (or very close to 0), while it does not provide information regarding entanglement, from the Schmidt decomposition it follows that the effective Hilbert space dimension of the subsystem gets reduced by one. 

In the next section we consider a system of coupled kicked tops and explore to what extent the behaviour of the smallest Schmidt eigenvalue is described by the fixed trace Wishart-Laguerre ensemble.
 
\section{Coupled kicked tops}
\label{CKT}

The kicked top system has been a paradigm for studying chaos, both classically and quantum mechanically~\cite{HKS1987,Haake2010}. Remarkably, it has also been realized experimentally using an ensemble of Caesium atoms~\cite{CSAGJ2009}. In the study of entanglement formation in bipartite systems, a coupled system of two kicked tops has turned out to be of great importance and has been investigated by a number of researchers~\cite{MS1999,BL2002,FMT2003,BL2004,TMD2008,KAT2013}.

The full Hamiltonian of the coupled kicked top system is
\begin{equation}
\label{CKTH}
H=H_1\otimes \1_{N_2}+\1_{N_1}\otimes H_2+H_{12}.
\end{equation}
Here, 
\begin{equation}
H_r=\frac{\pi}{2}J_{y_r}+\frac{k_r}{2j_r}J_{z_r}^2\sum_{\nu=-\infty}^{\infty}\delta(t-\nu),~~r=1,2,
\end{equation}
represent the Hamiltonians for the individual tops, and
\begin{equation}
H_{12}=\frac{\epsilon}{\sqrt{j_1j_2}}(J_{z_1}\otimes J_{z_2})\sum_{\nu=-\infty}^{\infty} \delta(t-\nu)
\end{equation}
is the interaction term. Also, $\1_{N_r}$ represents identity operator that acts on $N_r$-dimensional Hilbert space $\mathcal{H}^{(N_r)}$. The Hamiltonians $H_1$ and $H_2$ correspond respectively to $N_1 ~(=2j_1+1)$-dimensional, and $N_2 ~(=2j_2+1)$-dimensional Hilbert spaces $\mathcal{H}^{(N_1)}$ and $\mathcal{H}^{(N_2)}$, respectively. The Hamiltonian for the coupled kicked tops corresponds to an $N_1 N_2$-dimensional Hilbert space $\mathcal{H}^{(N_1N_2)}=\mathcal{H}^{(N_1)}\otimes\mathcal{H}^{(N_2)}$. Also, $(J_{x_r}, J_{y_r}, J_{y_r})$ are angular momentum operators for the $r$th top and satisfy the usual commutation relations. The parameter $k_r$ controls the chaotic behaviour of the individual tops. The parameter $\epsilon$ takes care of the coupling between the two tops.

The unitary time evolution operator (Floquet operator) corresponding to the Hamiltonian~(\ref{CKTH}) is
\begin{eqnarray}
U=(U_1\otimes U_2)U_{12},
\end{eqnarray}
with
\begin{equation}
U_r=\exp\left(-\frac{\iota\pi}{2}J_{y_r}-\frac{\iota k_r}{2j_r}J_{z_r}^2\right),r=1,2;
\end{equation}
\begin{equation}
U_{12}=\exp\left(-\frac{\iota\epsilon}{\sqrt{j_1j_2}}~J_{z_1}\otimes J_{z_2}\right).
\end{equation}
Here $\iota=\sqrt{-1}$ represents the imaginary unit.
The initial state for the individual tops is chosen as a generalized SU(2) coherent state or the directed angular momentum state~\cite{HKS1987,Haake2010}, which is given in $|j_r,m_r\rangle$ basis as 
$
\langle j_r,m_r|\theta_0^{(r)},\phi_0^{(r)}\rangle=\left(1+|\gamma_r|^2\right)^{-j_r} \gamma_r^{j_r-m_r}\sqrt{\left(2j_r \atop j_r+m_r\right)}
$
with $\gamma_r\equiv \exp(\iota\phi_0^{(r)})\tan(\theta_0^{(r)}/2)$. For later use, we define $N_r$-dimensional vectors given by
\begin{equation}
\boldsymbol{\chi}_{r}=[\langle j_r,m_r|\theta_0^{(r)},\phi_0^{(r)}\rangle]_{m_r=-j_r,\ldots,+j_r}.
\end{equation}

For the coupled top, the initial state is taken as the tensor-product of the states of the individual tops: $
|\psi(0)\rangle=|\theta_0^{(1)},\phi_0^{(1)}\rangle \otimes |\theta_0^{(2)},\phi_0^{(2)}\rangle$. We can implement the time evolution to obtain the state $|\psi(\nu)\rangle$ starting from $|\psi(0)\rangle$ using the iteration scheme
$
|\psi(\nu)\rangle=U|\psi(\nu-1)\rangle=(U_1\otimes U_2)U_{12}|\psi(\nu-1)\rangle,
$
which, when written in $\langle j_1,s_1; j_2,s_2|$ basis, is~\cite{MS1999}
\begin{eqnarray*}
&&\langle j_1,s_1; j_2,s_2| \psi(\nu)\rangle=\exp\left(-\iota\frac{\epsilon}{\sqrt{j_1j_2 }}s_1 s_2\right)\\
&&\times\!\sum_{m_1=-j_1}^{+j_1}\sum_{m_2=-j_2}^{+j_2}\langle j_1,s_1|U_1|j_1,m_1\rangle\langle j_2,s_2|U_2|j_2,m_2\rangle\langle j_1,m_1; j_2,m_2 | \psi(\nu-1)\rangle.
\end{eqnarray*}
\begin{figure*}[h!]
\centering
\includegraphics[width=01\textwidth]{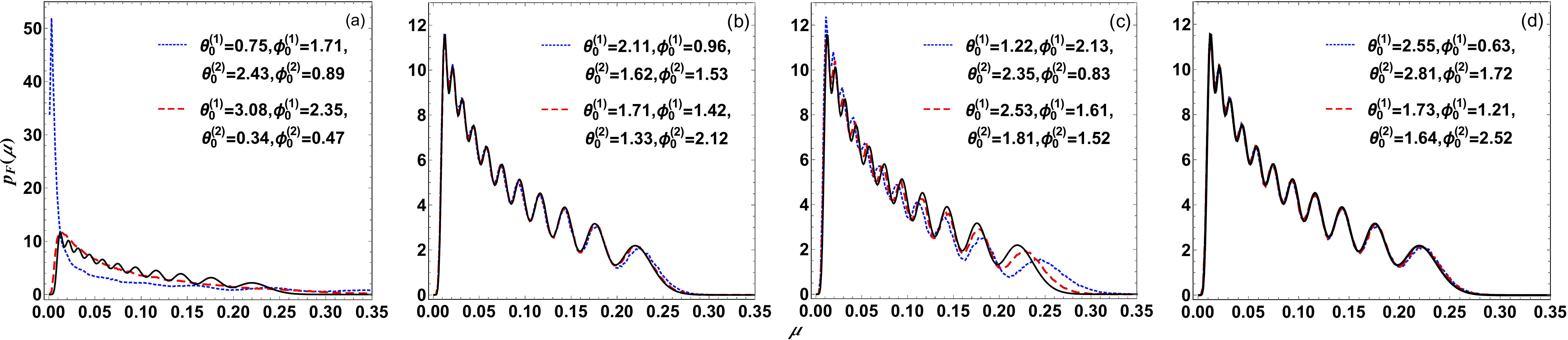}
\includegraphics[width=01\textwidth]{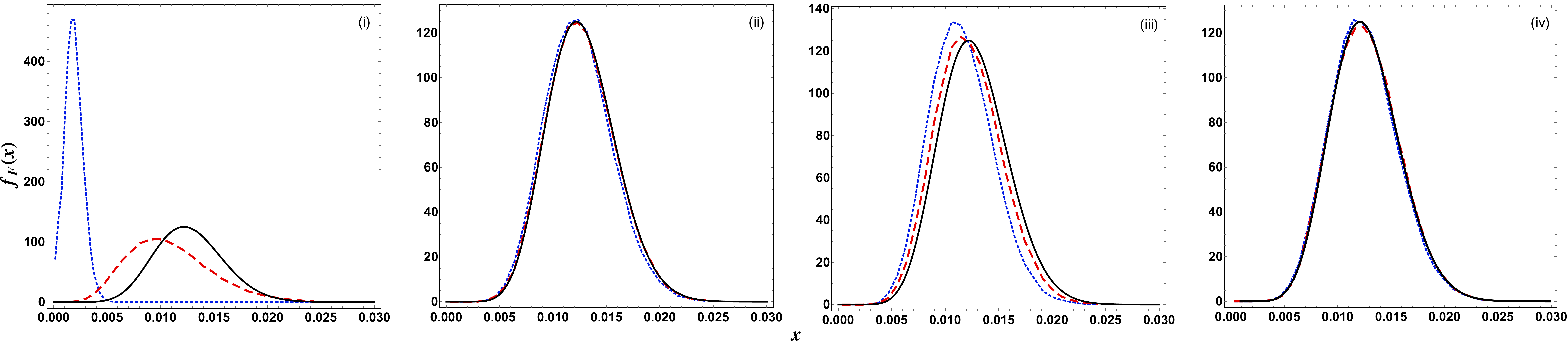}
\caption{\small Comparison of marginal density (top row) and the smallest eigenvalue density (bottom row) of the coupled kicked top system with the fixed trace ensemble results for $N_1=11,N_2=21$ and $\epsilon=1$. For plots (a)-(d) as well as (i)-(iv), the parameters ($k_1,k_2$) vary as (0.5, 1), (0.5, 8), (2.5, 3), (7, 8). In each case the solid (black) curves correspond to the analytical results, while the dotted (blue) and dashed (red) curves correspond to different initial conditions. The $\theta_0$ and $\phi_0$ values used for different initial conditions are mentioned in (a)-(d), and hold, respectively, for (i)-(iv) also.}
 \label{k_var}
\end{figure*}
 \begin{figure*}[t!]
\centering
\includegraphics[width=1\textwidth]{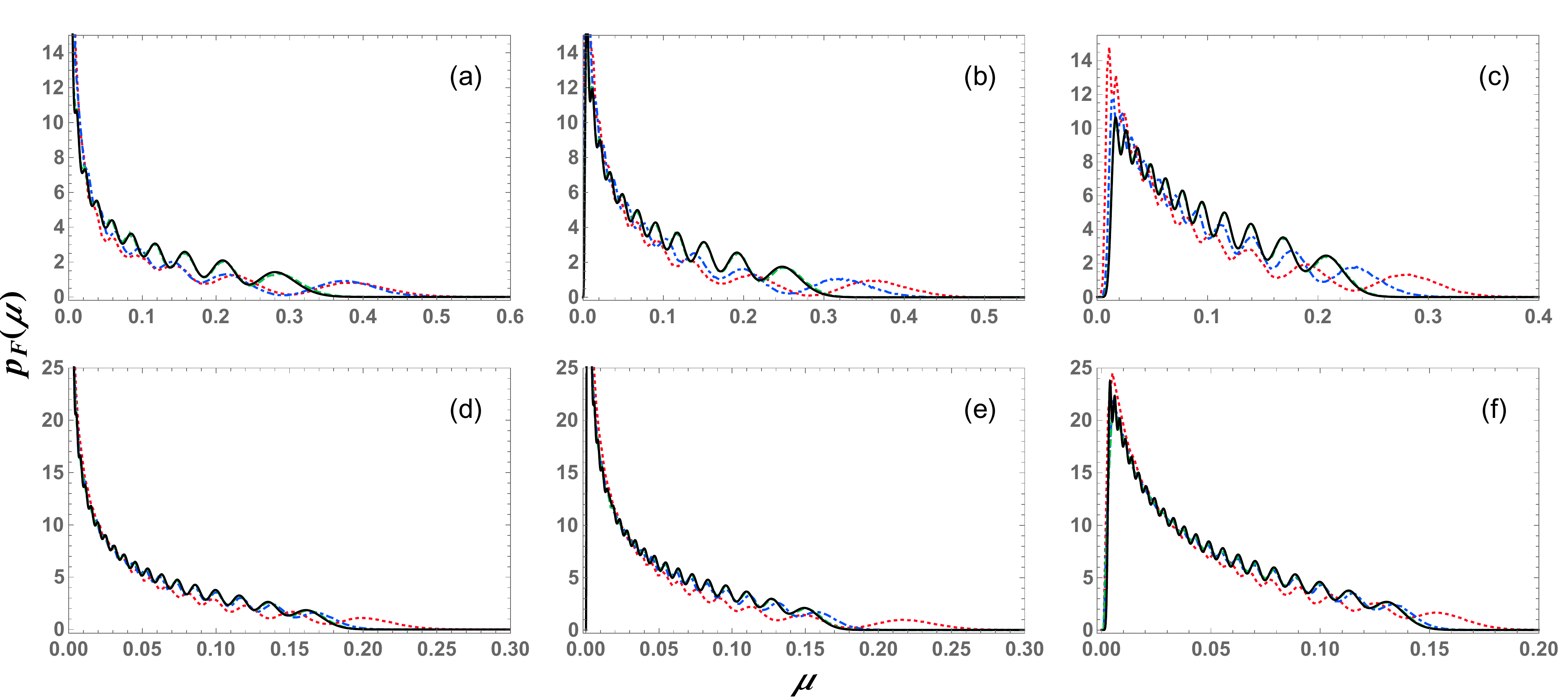}
\includegraphics[width=1\textwidth]{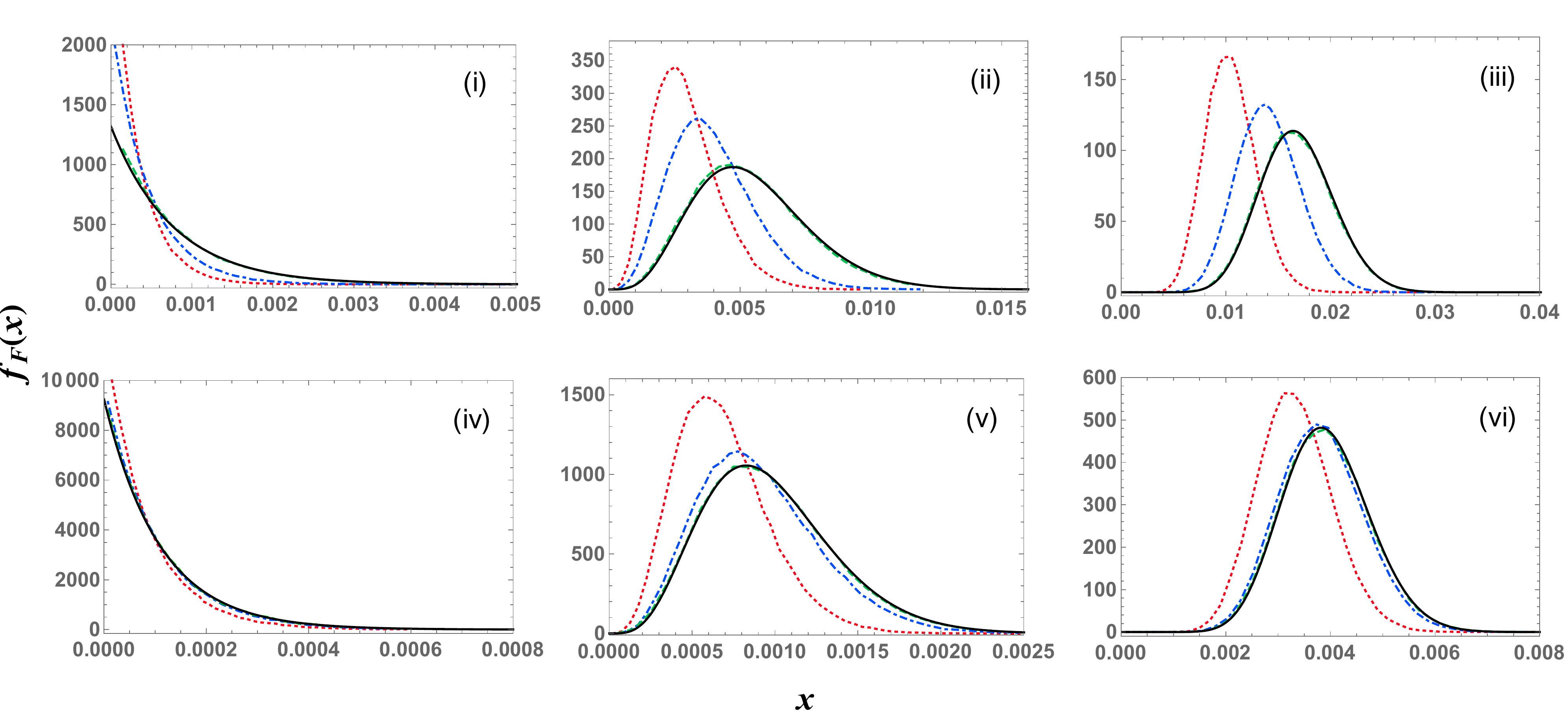}
\caption{\small Effect of varying $\epsilon$ on marginal density ((a)-(f)), and the smallest eigenvalue density ((i)-(vi)). The solid lines (black) correspond to analytical results, while the dotted (red), dot-dashed (blue) and dashed (green) curves result from coupled top simulation for $\epsilon=0.05, 0.1$ and 0.5, respectively. The parameters $k_1,k_2$ are fixed at 7, 8, while dimension parameters ($N_1,N_2$) vary for the figures (a)-(f) as well as (i)-(vi) as (11, 11), (11, 15), (11, 25), (21, 21), (21, 25), (21, 35).}
 \label{eps_var}
\end{figure*}

A convenient approach for implementing this iteration scheme and eventually calculating the reduced density matrix involves writing the states as $N_1 \times N_2$ matrices:
\begin{equation}
\label{evol}
\boldsymbol{\Psi}(\nu)={\bf V}\circ({\bf U}_1 \boldsymbol{\Psi}(\nu-1) {\bf U}_2^T ).
\end{equation}
Here `$\circ$' represents the Hadamard product and `$T$' the transpose. $\bf V$ is an $N_1\times N_2$ matrix given by
\begin{equation}
{\bf V}=\left[\exp\left(-\iota\frac{\epsilon}{\sqrt{j_1j_2 }}a\,b\right)\right]_{{{a=-j_1,\ldots,+j_1}}\atop{b=-j_2,\ldots,+j_2}}.
\end{equation}
 Also, ${\bf U}_r$ is an $N_r\times N_r$  dimensional matrix
\begin{eqnarray}
{\bf U}_r=\left[\exp\left(-\iota\frac{k_r}{2j_r}a^2\right) \,d_{a,b}^{(j_r)}\left(\frac{\pi}{2}\right)\right]_{a,b=-j_r,\ldots,+j_r}.
\end{eqnarray}
Here $d_{a,b}^{(j_r)}$ represents the Wigner (small) $d$ matrix elements. We use the inbuilt function in Mathematica~\cite{Mathematica} for Wigner (big) $D$ matrix to evaluate it. The initial $N_1\times N_2$-dimensional state matrix is given by
\begin{equation}
\boldsymbol{\Psi}(0)=\boldsymbol{\chi}_{1}\otimes\boldsymbol{\chi}_{2}^T.
\end{equation}
Eventually, the reduced density matrix of dimension $N_1\times N_1$ can be constructed as
\begin{equation}
\label{rhod}
\rho_d=\boldsymbol{\Psi}(\nu)\boldsymbol{\Psi}(\nu)^\dag.
\end{equation}
The eigenvalues of this matrix are the sought after Schmidt eigenvalues, whose statistics is of interest to us. To obtain an ensemble of states we proceed as follows. We begin with an initial state and apply~(\ref{evol}) iteratively. After ignoring initial 500 states to safely avoid the transient regime~\cite{BL2004}, we start considering states separated by 20 time steps to put off any unwanted correlations. In all, we consider 50000 states for statistical analysis. 

In figure~\ref{k_var} we set $N_1=11, N_2=21, \epsilon=1$, and examine the effect of different choices of $k_1,k_2$ on one level density and smallest eigenvalue density for the coupled kicked tops. For (a), (i) we have $k_1=0.5,k_2=1$ for which the classical phase spaces of the individual tops consist mostly of regular orbits~\cite{BL2004}. In this case, we can see deviations from the fixed trace ensemble results with strong sensitivity to initial conditions, i.e. $\theta^{(r)}_0$ and $\phi^{(r)}_0$ values. In (b), (ii) we set $k_1=0.5,k_2=8$. In this case highly chaotic phase space~\cite{BL2004} of the second top leads to an agreement with the results of the fixed trace ensemble, even though the phase space of the first top is mostly regular. Moreover, there is a weak sensitivity to initial conditions. In (c), (iii) we consider $k_1=2.5, k_2=3$, both of which correspond to mixed type phase space~\cite{BL2004}. Here we observe deviations, along with some sensitivity to initial conditions. Finally, in (d), (iv) we take $k_1=7,k_2=8$, for which phase spaces of both the tops are highly chaotic. In this case, we have very good agreement with the random matrix results and very weak sensitivity to the initial conditions.

In figure~\ref{eps_var} we consider a chaotic regime ($k_1=7, k_2=8$) and examine the effect of varying $\epsilon$ for various combinations of $n$ and $\alpha$. We observe that for a given $\epsilon$, increase in $n$ or $\alpha$ leads to a better agreement with the fixed trace ensemble results. Recent studies in a similar direction have investigated the universal aspects of spectral fluctuations and entanglement transitions in strongly chaotic subsystems~\cite{STLKB2016,LSKBT2016}.

A quantifier to measure the fraction of close to {\it maximally} entangled states can be the cumulative probability $R(\delta)=\int_{1/n-\delta}^{1/n} f_F(x)\,dx$~\cite{Majumdar2011}, that turns out to be vanishingly small for $\delta<<1/n$ and thus, implies that the actual fraction of such states is extremely small. For example, using~(\ref{fFW}), we obtain $R(\delta=0.1/ n) $ value to be roughly (i) $8\times 10^{-6}$ for $n=3, m=11$, (ii) $1\times 10^{-35}$ for $n=7, m=19$, and (iii)  $5\times 10^{-91}$ for $n=11, m=25$.

\section{Summary and conclusion}
\label{SAC}
We considered complex Wishart-Laguerre ensemble, both with and without the fixed trace condition, and provided an easily implementable recurrence scheme to obtain the exact result for the density of the smallest eigenvalue. This method also gives access to arbitrary moments of the smallest eigenvalue. The recursion-based approach for exact and explicit expressions for the density is preferable to the results based on determinants which are difficult to handle with increasing dimensionality $n$ or rectangularity $\alpha$. We also demonstrated the equivalence of the recurrence scheme and the determinant-based results for $\alpha=0$ and 1. We validated our analytical results using Monte Carlo simulations and also used large $n$ and $\alpha$ evaluations to compare with the Tracy-Widom density. As an application to quantum entanglement problem we explored the behaviour of Schmidt eigenvalues of the coupled kicked top system. Among other things, we found that in the chaotic regime, the fixed trace ensemble describes the behaviour of the Schmidt eigenvalues very well if sufficient coupling is provided between the constituent tops. 

\ack This work initiated from a project that was carried out at Shiv Nadar University under the Opportunities for Undergraduate Research (OUR) scheme. The authors are grateful to the anonymous reviewers for fruitful suggestions.

\appendix

\section{Recurrence scheme}
\label{AppA}

We begin with (\ref{fWdef}) and apply the shift $\lambda_i\rightarrow \lambda_i+x$. This results in
\begin{equation}
\fl
f(x)=nC_{n,\alpha}\, x^\alpha e^{-n x} \int_0^\infty d\lambda_2\cdots \int_0^\infty d\lambda_n\, \prod_{2\leq k<j\leq n}(\lambda_j-\lambda_k)^2\prod_{i=2}^{n}\lambda_i^2(\lambda_i+x)^\alpha e^{-\lambda_i}.
\end{equation}
To derive the recurrence relation, we will proceed parallel to the steps in chapter 4 of~\cite{Edelman1989}, or as in~\cite{Edelman1991}. To this end, we shift the indices of the integration variables as $\lambda_i\rightarrow \lambda_{i-1}$, and also introduce the measure $d\Omega_i=\lambda_i^2 \,e^{-\lambda_i}\,d\lambda_i$. Consequently, we arrive at the following expression:
\begin{eqnarray}
f(x)=nC_{n,\alpha}\, x^\alpha e^{-n x} \int_0^\infty d\Omega_1\cdots \int_0^\infty d\Omega_{n-1}\Delta_{n-1}^2(\{\lambda\})\prod_{i=1}^{n-1}(\lambda_i+x)^\alpha.
\end{eqnarray}
Next, we define
\begin{equation}
\label{def1}
I_{i,j}^\alpha=\int_0^\infty d\Omega_1\cdots\int_0^\infty d\Omega_{n-1}\, \Delta_{n-1}^2(\{\lambda\})u(x),
\end{equation} 
where the integrand $u(x)$ is
\begin{eqnarray}
\nonumber
&&\underbrace{(\lambda_1+x)^\alpha\cdots(\lambda_i+x)^\alpha}_{i \mathrm{~ terms~}}\underbrace{(\lambda_{i+1}+x)^{\alpha-1}\cdots(\lambda_{i+j}+x)^{\alpha-1}}_{j \mathrm{ ~terms~}}\\
&&\times \underbrace{(\lambda_{i+j+1}+x)^{\alpha-2}\cdots(\lambda_n+x)^{\alpha-2}}_{n-i-j-1 \mathrm{~ terms~}}.
\end{eqnarray}
We also define the operator
\begin{eqnarray}\label{def2}
I_{i,j}^\alpha[v]=\int_0^\infty d\Omega_1\cdots\int_0^\infty d\Omega_{n-1} \,\Delta_{n-1}^2(\{\lambda\})u(x)\,v. 
\end{eqnarray}
Using the above notation the smallest eigenvalue density can be written as
\begin{equation}
\label{sev}
f(x)=nC_{n,\alpha} x^\alpha e^{-nx}I_{n-1,0}^\alpha.
\end{equation}
With these, Lemma 4.2 of~\cite{Edelman1989} (or, equivalently, Lemma 4.1 of~\cite{Edelman1991}) holds in the complex case also:
\begin{equation}
\label{op_result}
I_{i,j}^\alpha[\lambda_k]=\cases{ I_{i+1,j-1}^\alpha-x I_{i,j}^\alpha &  if  $i<k\leq i+j$,\\
                                  I_{i,j+1}^\alpha-x I_{i,j}^\alpha & if  $i+j < k< n$.}
\end{equation}
The above result follows by writing $\lambda_k$ as $(\lambda_k+x)-x$ and then using the operator defined in~(\ref{def2}). 
Next, if the terms $(\lambda_k+x)$ and $(\lambda_l+x)$ share the same exponent in the integrals (i.e., both $k$ and $l$ fall within one of the closed intervals $[1,i],[i+1,i+j]$, or $[i+j+1,n-1]$), then
\begin{eqnarray}
\label{eqa}
I_{i,j}^\alpha\left[\frac{\lambda_k\lambda_l}{\lambda_k-\lambda_l}\right]=0,\\
\label{eqb}
I_{i,j}^\alpha\left[\frac{\lambda_k}{\lambda_k-\lambda_l}\right]=\frac{1}{2}I_{i,j}^\alpha,\\
\label{eqc}
I_{i,j}^\alpha\left[\frac{\lambda_k^2}{\lambda_k-\lambda_l}\right]=I_{i,j}^{\alpha}[\lambda_k].
\end{eqnarray}
Equation~(\ref{eqa}) follows because of the asymmetry in $\lambda_k$ and $\lambda_l$, while ~(\ref{eqb}) is obtained using the identity $\lambda_k/(\lambda_k-\lambda_l)+\lambda_l/(\lambda_l-\lambda_k)=1$ and using symmetry. The result~(\ref{eqc}) is obtained using the identity $\lambda_k^2/(\lambda_k-\lambda_l)=\lambda_k+\lambda_k\lambda_l/(\lambda_k-\lambda_l)$ and (\ref{eqa}). 

The crucial difference occurs in the first equation of Lemma 4.3~\cite{Edelman1989} (or Lemma 4.2~\cite{Edelman1991}), which reads for the present case as
\begin{eqnarray}
\fl
&&I_{i,j}^\alpha=(x+\alpha+j+2k+2)I_{i-1,j+1}^\alpha-x[k+(\alpha-1)]I_{i-1,j}^\alpha+(i-1)x I_{i-2,j+2}^\alpha \label{Rec1}\\
\fl
&&I_{0,j}^\alpha=I_{j,n-j-1}^{\alpha-1},\label{Rec2}
\end{eqnarray}
with $k=n-i-j-1$. The second equation of this set, (\ref{Rec2}), follows readily from the definition~(\ref{def1}). This first equation of this set, (\ref{Rec1}), is derived using 
\begin{equation}
I_{i,j}^\alpha=x I_{i-1,j+1}^\alpha+I_{i-1,j+1}^\alpha[\lambda_i],
\end{equation}
which is a consequence of~(\ref{op_result}). The difference in the result compared to the real case occurs due to the term $I_{i-1,j+1}^\alpha[\lambda_i]$. For the complex case, this involves observing the following:
\begin{equation}
\fl
\int_0^\infty (\lambda_i+x)^{\alpha-1}\prod_{i<l}(\lambda_l-\lambda_i)^2\,\lambda_i^3\,e^{-\lambda_i}\,d\lambda_i=\int_0^\infty  \frac{d}{d\lambda_i} [(\lambda_i+x)^{\alpha-1}\prod_{i<l}(\lambda_l-\lambda_i)^2\,\lambda_i^3]e^{-\lambda_i}\,d\lambda_i.
\end{equation}
Also, using the result 
\begin{equation}
\frac{dI_{i-1,j+1}^\alpha}{dx}=(i-1)\alpha I_{i-2,j+2}^\alpha+(j+1)(\alpha-1)I_{i-1,j}^\alpha
\end{equation}
for $i+j=n-1$, as given in the proof for Lemma 4.4 of~\cite{Edelman1989} (or Lemma 4.3 of~\cite{Edelman1991}), yields in the present case
\begin{eqnarray}
\label{Rec3}
\fl
I_{i,j}^\alpha=(x+\alpha+j+2)I_{i-1,j+1}^\alpha-\frac{x}{j+1}\frac{d}{dx}I_{i-1,j+1}^\alpha+x (i-1)\left(1+\frac{\alpha}{j+1}\right)I_{i-2,j+2}^\alpha.
\end{eqnarray}
Now, we can begin with $I_{n-1,0}^{\alpha-1}$, which is same as $I_{0,n-1}^\alpha$ in view of~(\ref{Rec2}). Equation~(\ref{Rec3}) can be used with $j=n-i-1$ repeatedly for $i=1$ to $n-1$ to arrive at $I_{n-1,0}^\alpha$, starting from $I_{0,n-1}^\alpha$. We note that $I_{n-1,0}^\alpha$ is the term needed to obtain the smallest eigenvalue density expression~(\ref{sev}) explicitly. This is essentially what has been employed in the recurrence involving $S_i:= I_{i,n-i-1}^\alpha/I_{n-1,0}^0$ for $g_{n,m}(x)$ in~(\ref{fW}). We also note that $I_{n-1,0}^0=1/C_{n-1,2}$. The constant $c_{n,m}$ of~(\ref{fW}) is therefore $nC_{n,\alpha}/C_{n-1,2}$.

\section{Mathematica codes}
\label{Codes}
The following code can be implemented in Mathematica~[28] to obtain exact expressions for the smallest eigenvalue density for the Wishart-Laguerre ensemble:
\begin{figure}[!h]
\centering
\includegraphics[width=0.85\linewidth]{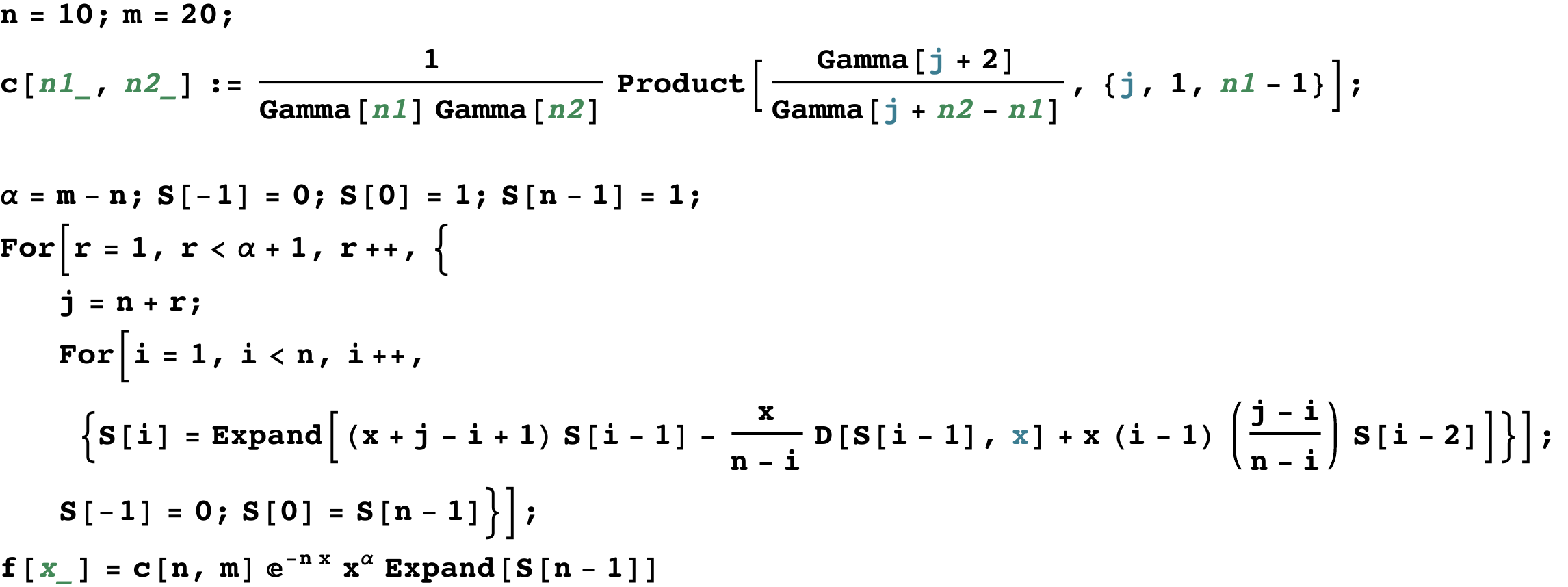}
\end{figure}

\pagebreak

For generating the smallest eigenvalue density for the fixed trace Wishart-Laguerre ensemble, the following code can be used along with the above.
\begin{figure}[!h]
\begin{flushleft}
~~ \includegraphics[width=0.95\linewidth]{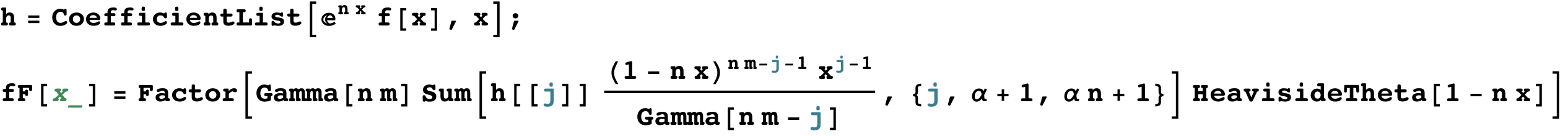}
\end{flushleft}
\end{figure}

\noindent
We can also directly implement the inverse Laplace transform function built in Mathematica:
\begin{figure}[!h]
\begin{flushleft}
~~\includegraphics[width=0.7\linewidth]{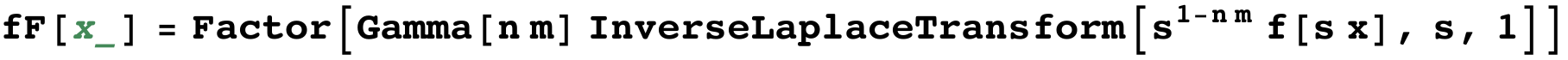}
\end{flushleft}
\end{figure}

\noindent
The `{\sf Factor}' option in the above codes is for printing compact expressions on-screen. For computation involving large $n$ or $\alpha$ values, it may be removed, since factoring very large expressions may result in a large computation time.

The moments of the smallest eigenvalue of the regular or the fixed-trace Wishart-Laguerre ensemble can be obtained using the following functions:
\begin{figure}[!h]
\begin{flushleft}
~~\includegraphics[width=0.52\linewidth]{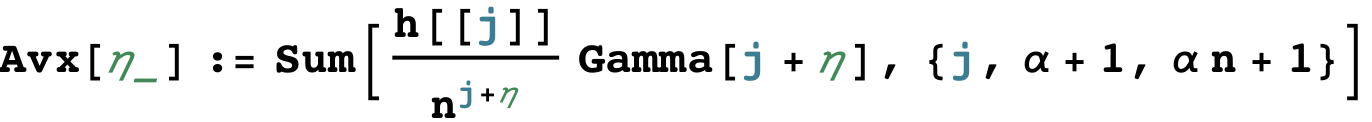}\\~\\
~~\includegraphics[width=0.31\linewidth]{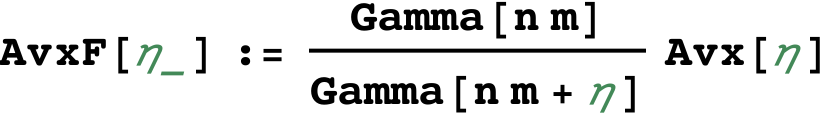}
\end{flushleft}
\end{figure}

\vspace{-0.8cm}

\section{Relation with associated Laguerre polynomial}
\label{EqProof}

The associated Laguerre polynomials satisfy the following relations~\cite{Szego1975}:
\begin{eqnarray}
i L_i^{(k)}(-x)=(x+k+1)L_{i-1}^{(k+1)}(-x)+x L_{i-2}^{(k+2)}(-x),\\
\frac{d}{dx}L_i^{(k)}(-x)=L_{n-1}^{(k+1)}(-x).
\end{eqnarray}
These two can be combined to obtain the following relation:
\begin{equation}
\fl
i L_i^{(k)}(-x)=(x+k+1)L_{i-1}^{(k+1)}(-x)-\frac{x}{k-1}\frac{d}{dx}L_{i-1}^{(k+1)}(-x)+\frac{x k}{k-1}L_{i-2}^{(k+2)}(-x).
\end{equation}
Considering $k=n-i+1$ gives
\begin{eqnarray}
\nonumber
i L_i^{(n-i+1)}(-x)=(x+n-i+2)L_{i-1}^{(n-i+2)}(-x)-\frac{x}{n-i}\frac{d}{dx}L_{i-1}^{(n-i+2)}(-x)\\
~~~~~~~~~~~~~~~~~~~~+\frac{x (n-i+1)}{n-i}L_{i-2}^{(n-i+3)}(-x).
\end{eqnarray}
Multiplying this equation by $\Gamma(i)$ and then calling $S_i=\Gamma(i+1)L_i(n-i+1)(-x)$, we get
\begin{eqnarray}
S_i=(n-i+2+x)S_{i-1}-\frac{x}{n-i}\frac{dS_{i-1}}{dx}+x(i-1)\frac{  (n-i+1)}{n-i}S_{i-2}.
\end{eqnarray}
This recurrence relation is the same as that given in section~\ref{Sec2} when used for $m=n+1$. Hence, $g_{n,n+1}(x)=S_{n-1}=\Gamma(n)L_{n-1}^{(2)}(-x)$. 

\section{Some explicit results}
\label{ERes}

For $\alpha=m-n=0$, the smallest eigenvalue density expressions valid for all $n$ are quite compact and are already provided in (\ref{fa0}) and (\ref{fFa0}), respectively, for the regular Wishart-Laguerre ensemble and for the fixed trace Wishart-Laguerre ensemble. For a few other cases we tabulate the exact results in Tables D1 and D2 using the above Mathematica codes. This includes the $\alpha=1$ case for which closed-form results for any $n$ are given in (\ref{fa1}) and (\ref{fFa1}). In the case of fixed trace ensemble there is a $\Theta(1-nx)$ term in each of the probability density expressions that we have not shown in the table for the sake of compactness.

\begin{table}[h!]
\scriptsize
\centering
\begin{tabular}{ |c|c|c| }
\hline
$n$ & $m$ & $f(x)$ \\ \hline \hline
\multirow{4}{*}{2} 
& 3 & $e^{-2 x} x \left(x+3\right)$ \\ \cline{2-3}
 & 4 & $e^{-2 x} x^2 \big(x^2+6 x+12\big)\big/6$ \\ \cline{2-3}
 & 5 & $ e^{-2 x} x^3  \big(x^3+9 x^2+36 x+60\big)/72$ \\ \cline{2-3}
 & 6 & $e^{-2 x} x^4  \big(x^4+12 x^3+72 x^2+240 x+360 \big)/1440$ \\ \hline\hline
\multirow{4}{*}{3} 
& 4 & $e^{-3 x} x  \big(x^2+8 x+12\big)\big/2$ \\ \cline{2-3}
 & 5 & $ e^{-3 x} x^2  \big(x^4+16 x^3+96 x^2+240 x+240\big)\big/48$ \\ \cline{2-3}
  & 6 & $e^{-3 x} x^3 \big(x^6+24 x^5+252 x^4+1440 x^3+4680 x^2+8640 x+7200\big)\big/2880$ \\ \cline{2-3}
 & 7 & $e^{-3 x} x^4 \big(x^8+32 x^7+480 x^6+4320 x^5+25200 x^4+97920 x^3+253440 x^2+403200 x+302400\big)\big/345600$ \\ \hline\hline
\multirow{4}{*}{4} 
& 5 & $e^{-4 x} x  \big(x^3+15 x^2+60 x+60\big)\big/6$ \\ \cline{2-3}
 & 6 & $e^{-4 x} x^2  \big(x^6+30 x^5+360 x^4+2160 x^3+6840 x^2+10800 x+7200\big)\big/720$ \\  \cline{2-3}
 & 7 & \makecell{$e^{-4 x} x^3  \big(x^9+45 x^8+900 x^7+10380 x^6+75600 x^5+360720 x^4+1130400 x^3+2268000 x^2$\\$+2721600 x+1512000\big)\big/259200$} \\  \cline{2-3}
 & 8 & \makecell{$e^{-4 x} x^4 \big (x^{12}+60 x^{11}+1680 x^{10}+28800 x^9+334800 x^8+2773440 x^7+16790400 x^6+74995200 x^5$\\$+246456000 x^4+586656000 x^3+972518400 x^2+1016064000 x+508032000\big)\big/217728000$} \\ \hline\hline
\multirow{4}{*}{5} 
& 6 & $e^{-5 x} x \big (x^4+24 x^3+180 x^2+480 x+360\big)\big/24$ \\ \cline{2-3}
 & 7 & $e^{-5 x} x^2 \big (x^8+48 x^7+960 x^6+10320 x^5+64800 x^4+241920 x^3+524160 x^2+604800 x+302400\big)\big/17280$ \\ \cline{2-3}
 & 8 & \makecell{$e^{-5 x} x^3 \big (x^{12}+72 x^{11}+2340 x^{10}+45120 x^9+572400 x^8+5019840 x^7+31157280 x^6+137894400 x^5$\\$+432734400 x^4+943488000 x^3+1371686400 x^2+1219276800 x+508032000\big)\big/43545600$} \\ \cline{2-3}
 & 9 & \makecell{$e^{-5 x} x^4  \big(x^{16}+96 x^{15}+4320 x^{14}+120480 x^{13}+2323440 x^{12}+32780160 x^{11}+349493760 x^{10}+2870380800 x^9$ \\ $+18353563200 x^8+91755417600 x^7+358177075200 x^6+1083937075200 x^5+2506629888000 x^4$\\$+4316239872000 x^3+5267275776000 x^2+4096770048000 x+1536288768000\big)\big/292626432000$} \\ \hline
\end{tabular}
\caption{Results for the Wishart-Laguerre ensemble} 
\end{table}


\begin{table}
\scriptsize
\begin{tabular}{ |c|c|c| }
\hline
$n$ & $m$ & $f_F(x)$ \\ \hline \hline
\multirow{4}{*}{2} 
& 3 & $60  x \big(1-x\big) \big(1-2x\big)^2 $ \\ \cline{2-3}
 & 4 & $420 x^2  \big(1-x\big)^2  \big(1-2 x\big)^2$ \\ \cline{2-3}
 & 5 & $2520 x^3 \big(1-x\big)^3 \big(1-2x\big)^2$ \\ \cline{2-3}
 & 6 & $13860 x^4  \big(1-x\big)^4  \big(1-2x\big)^2$ \\ \hline\hline
\multirow{4}{*}{3} 
& 4 & $660 x  \big(1-3 x^2\big)\big(1-3 x\big)^7$ \\ \cline{2-3}
 & 5 & $10920 x^2 \big(1-x-x^2-9 x^3+15 x^4\big)\big(1-3 x\big)^7$ \\ \cline{2-3}
  & 6 & $28560 x^3  \big(5-12 x+12 x^2-48 x^3-48 x^4+432 x^5-411 x^6\big)\big(1-3 x\big)^7$ \\ \cline{2-3}
 & 7 & $1627920 x^4 \big(1-4 x+8 x^2-16 x^3+320 x^6-756 x^7+489 x^8\big)\big(1-3x\big)^7$ \\ \hline\hline
\multirow{4}{*}{4} 
& 5 & $3420 x  \big(1+5 x-20 x^2+4 x^3\big)\big(1-4x)^{14}$ \\ \cline{2-3}
 & 6 & $106260 x^2   \big(1 + 6 x + x^2 - 204 x^3 + 486 x^4 - 424 x^5 + 356 x^6\big)\big(1 - 4 x\big)^{14}$ \\  \cline{2-3}
 & 7 & \makecell{$491400 x^3  \big(5+27 x+51 x^2-683 x^3-5286 x^4+35910 x^5-85295 x^6+116895 x^7$\\$-79980 x^8-9196 x^9\big)\big(1-4x\big)^{14}$ }\\  \cline{2-3}
 & 8 & \makecell{$6796440 x^4 \big(7+28 x+86 x^2-540 x^3-6775 x^4-18416 x^5+440876 x^6-2012008 x^7$\\$+4901710 x^8-7145600 x^9+5855692 x^{10}-3288592 x^{11}+2386196 x^{12}\big)\big(1-4x\big)^{14}$} \\ \hline\hline
\multirow{4}{*}{5} 
& 6 & $12180 x \big(1+16 x-39 x^2-140 x^3+220 x^4\big) \big(1-5 x\big)^{23} $ \\ \cline{2-3}
 & 7 & $628320 x^2  \big(1+22 x+142 x^2-1234 x^3-580 x^4+4676 x^5+29788 x^6-92420 x^7+75355 x^8\big) \big(1-5x\big)^{23} $ \\ \cline{2-3}
 & 8 & \makecell{$23030280 x^3\big(1+24 x+243 x^2+280 x^3-19962 x^4+50208 x^5-31022 x^6+649056 x^7-1420095 x^8$\\$-7867032 x^9+35763831 x^{10}-53675640 x^{11}+27627140 x^{12}\big) \big(1-5x\big)^{23} $} \\ \cline{2-3}
 & 9 & \makecell{$97740720 x^4  \big(7+168 x+1968 x^2+9642 x^3-75517 x^4-1457898 x^5+10143328 x^6-31939648 x^7$\\$+134132583 x^8-323536148 x^9-511260568 x^{10}+786421818 x^{11}+22191959881 x^{12}$\\$-105911938466 x^{13}+211492028376 x^{14}-203837200540 x^{15}+80216630930 x^{16}\big) \big(1-5x\big)^{23} $} \\ \hline
\end{tabular}
\caption{Results for the fixed trace Wishart-Laguerre ensemble} 
\end{table}

\clearpage


\end{document}